\documentclass[superscriptaddress,
preprint,
prd,tightenlines,showpacs,nofootinbib,eqsecnum,
amsfonts,amsmath,amssymb]{revtex4}

\usepackage{bm}
\usepackage{graphicx} 
\usepackage{hyperref}
\usepackage{mathrsfs}
\usepackage{multirow}

\newcommand{\ud}{\mathrm{d}}

\newcommand{\calO}{\mathcal{O}}

\usepackage{ulem}
\usepackage[usenames]{color}

\allowdisplaybreaks

\begin{document}

\title{High-order half-integral conservative post-Newtonian
  coefficients\\in the redshift factor of black hole binaries}

\author{Luc Blanchet}\email{blanchet@iap.fr}
\affiliation{$\mathcal{G}\mathbb{R}\varepsilon{\mathbb{C}}\mathcal{O}$,
  Institut d'Astrophysique de Paris --- UMR 7095 du CNRS,
  \\ Universit\'e Pierre \& Marie Curie, 98\textsuperscript{bis}
  boulevard Arago, 75014 Paris, France}

\author{Guillaume Faye}\email{faye@iap.fr}
\affiliation{$\mathcal{G}\mathbb{R}\varepsilon{\mathbb{C}}\mathcal{O}$,
  Institut d'Astrophysique de Paris --- UMR 7095 du CNRS,
  \\ Universit\'e Pierre \& Marie Curie, 98\textsuperscript{bis}
  boulevard Arago, 75014 Paris, France}

\author{Bernard F. Whiting}\email{bernard@phys.ufl.edu}
\affiliation{Institute for Fundamental Theory, Department of Physics,
  University of Florida, Gainesville, FL 32611, USA}
\affiliation{$\mathcal{G}\mathbb{R}\varepsilon{\mathbb{C}}\mathcal{O}$,
  Institut d'Astrophysique de Paris --- UMR 7095 du CNRS,
  \\ Universit\'e Pierre \& Marie Curie, 98\textsuperscript{bis}
  boulevard Arago, 75014 Paris, France}

\date{\today}

\begin{abstract}
The post-Newtonian approximation is still the most widely used
approach to obtaining explicit solutions in general relativity,
especially for the relativistic two-body problem with arbitrary mass
ratio. Within many of its applications, it is often required to use a
regularization procedure. Though frequently misunderstood, the
regularization is essential for waveform generation without reference
to the internal structure of orbiting bodies. In recent years, direct
comparison with the self-force approach, constructed specifically for
highly relativistic particles in the extreme mass ratio limit, has
enabled preliminary confirmation of the foundations of both
computational methods, including their very independent regularization
procedures, with high numerical precision. In this paper, we build
upon earlier work to carry this comparison still further, by examining
next-to-next-to-leading order contributions beyond the half integral
5.5PN conservative effect, which arise from terms to cubic and higher
orders in the metric and its multipole moments, thus extending
scrutiny of the post-Newtonian methods to one of the highest orders
yet achieved. We do this by explicitly constructing tail-of-tail terms
at 6.5PN and 7.5PN order, computing the redshift factor for compact
binaries in the small mass ratio limit, and comparing directly with
numerically and analytically computed terms in the self-force
approach, obtained using solutions for metric perturbations in the
Schwarzschild space-time, and a combination of exact series
representations possibly with more typical PN expansions. While self-force
results may be relativistic but with restricted mass ratio, our 
methods, valid primarily in the weak-field slowly-moving regime, are
nevertheless in principle applicable for arbitrary mass ratios.
\end{abstract}


\maketitle


\section{Introduction}

Over the last five years, comparison between post-Newtonian and gravitational
self-force calculations has made rapid progress, in large part due to both
high precision numerical computations from a self-force
perspective~\cite{Det08, BarackS09, BDLW10a, BDLW10b, Keidl10, Shah11, Shah12,
  Akcay12} (either by directly linearizing the Einstein field equations or by
using the Teukolsky equation~\cite{Teukolsky72, Teukolsky73, Keidl07}
or the Regge-Wheeler and Zerilli equations~\cite{ReggeW57, Ze70}), and
extensive analytical computations within the post-Newtonian
approximation~\cite{BDLW10a, BDLW10b, LBW12}. Much more recently, the
possibility for this comparison has been dramatically extended. From the
self-force side~\cite{SFW14, BiniD13, BiniD14a, BiniD14b}, this is due to the
new application of (already more than 15 years old) techniques~\cite{MST96a,
  MST96b, MT97} with which to represent metric perturbation solutions for
black hole space-times. On the post-Newtonian side, this has required the
computation of previously unevaluated higher order terms including
tail-of-tail effects~\cite{BFW14a} and, in particular, half-integral
$\frac{n}{2}$PN terms that are nevertheless conservative. In this paper, we
extend that most recent work. As will be seen, although the computations are
indeed very extensive, the results are quite simple to state and along with
further motivation, they are listed below, before we describe in detail the
processes we have used in their derivation.

\subsection{Motivation}

The self-force problem concerns itself with computations for binary
orbiting systems composed of compact bodies in which the mass ratio is
extreme, such that a full numerical relativity approach is unfeasible,
due to the vastly different length scales associated with the very
different masses and physical sizes of the compact bodies.
Foundations for the gravitational self-force (GSF) computations of
compact binaries have developed over the last two
decades~\cite{MiSaTa, QuWa, DW03, GW08, Pound10} (see
Refs.~\cite{PoissonLR, Detweilerorleans, Barackorleans} for reviews),
following very early work by De Witt and Brehme more than half a
century ago~\cite{dWB60}.  For the conservative part of the dynamics,
this has led to the recent possibility of high-order comparisons
between self-force computations~\cite{Det08, BDLW10a, BDLW10b} on the
one hand, and traditional post-Newtonian calculations (reviewed in
Ref.~\cite{Bliving14}) on the other hand, with ever increasing
precision.

For compact binaries moving on exactly circular orbits, Detweiler~\cite{Det08}
introduced a gauge invariant redshift factor, computed it numerically, and
showed agreement with existing post-Newtonian (PN) analytical
calculations~\cite{BFP98} up to 2PN order. Then a systematic program of
comparison was initiated in Refs.~\cite{BDLW10a, BDLW10b} which showed that
GSF and PN methods agree for the 3PN term and specific logarithmic
tail-induced contributions arising at 4PN and 5PN orders, and predicted
numerically the values of high-order PN coefficients, notably the full 4PN
coefficient. The analytical 4PN coefficient was then obtained~\cite{BiniD13}
using a combination of analytical self-force (SF) computation and a partial
derivation of the 4PN equations of motion in the Arnowitt-Deser-Misner (ADM)
Hamiltonian formalism~\cite{JaraS12, JaraS13}, with very good agreement with
the numerical value computed in Ref.~\cite{BDLW10b}.

Since that work, the accuracy of the numerical computation of the GSF has
improved drastically~\cite{SFW14}. The PN coefficients of the redshift factor
were obtained numerically to 10.5PN order and for a subset of coefficients,
also analytically, specifically those that are either rational, or made of the
product of $\pi$ with a rational, or a simple sum of commonly occurring
transcendentals~\cite{SFW14}. An alternative self-force
approach~\cite{BiniD14a, BiniD14b} (based on the post-Minkowskian expansion of
the Regge-Wheeler-Zerilli (RWZ) equation following Refs.~\cite{MST96a, MST96b,
  MT97}) has also obtained high order PN coefficients analytically, up to
8.5PN order.

A feature of the post-Newtonian expansion at high order is the
appearance of \textit{half-integral} PN coefficients (of type
$\frac{n}{2}$PN where $n$ is an odd integer) in the conservative
dynamics of binary point particles, moving on exactly circular
orbits. Using standard post-Newtonian methods it was
proved~\cite{BFW14a}\footnote{Hereafter we refer to this paper as
  Paper~I.} that the dominant half-integral PN term occurs at the
5.5PN order (confirming the finding of Ref.~\cite{SFW14}) and
originates from the non-linear ``tail-of-tail''
integrals~\cite{B98tail}. Here we continue Paper~I and compute, still
using the traditional PN method (in principle applicable for any mass
ratio), high-order half-integral PN terms at orders 6.5PN and 7.5PN in
the redshift factor, thus corresponding to the next-to-next-to-leading
half-integral contributions.

\subsection{Results}

We have computed the redshift factor introduced in
Ref.~\cite{Det08}, for a particle moving on an exact circular orbit
around a Schwarzschild black hole. The ensuing space-time is helically
symmetric, with a helical Killing vector $K^\alpha$ such that its
value $K_1^\alpha$ at the location of the particle is proportional to
the normalized four-velocity $u_1^\alpha$ of the particle,
\begin{equation}\label{utdef}
   u_1^\alpha = u_1^T \,K_1^\alpha\,.
\end{equation}
The redshift factor, denoted $u_1^T$, is thus defined geometrically as
the conserved quantity associated with the helical Killing symmetry
appropriate to conservative space-times with circular orbits. However,
adopting a coordinate system in which the helical Killing vector reads
$K^\alpha\partial_ \alpha = \partial_t + \Omega\,\partial_\varphi$,
where $\Omega$ is the orbital frequency of the circular motion, the
redshift factor reduces to the $t$ component $\ud t/\ud\tau_1$ of the
particle's four-velocity (where $\ud\tau_1$ is the particle's proper
time), and is thereby obtained as
\begin{equation}\label{uT}
    u_1^T = \biggl[- g_{\alpha\beta}(y_1) \frac{v_1^\alpha
        v_1^\beta}{c^2} \biggr]^{-1/2} \,,
\end{equation}
where $g_{\alpha\beta}(y_1)$ is the regularized metric evaluated at the particle's
location $y_1^\alpha = (c t, y_1^i)$, which we shall compute in detail
in the present paper for insertion into the redshift
factor~\eqref{uT}, and where $v_1^\alpha=\ud y_1^\alpha/\ud t = (c,
v_1^i)$ is the coordinate velocity.

In a first stage, our calculation is valid for a general extended
matter source, in the vacuum region outside the source. Then, in a
second stage, we use a matching argument to continue that solution
inside the source, which is then specialized to a binary point
particle system. Finally the metric is evaluated at the location of
one of the particles, with the help of a self-field regularization, in
principle dimensional regularization. Using the relative frame of the
center of mass and reducing the expressions to circular orbits,
mindful of the modification of the relation between the orbital
separation and the orbital frequency, we finally obtain the redshift
factor in the limit of a small mass ratio $q=m_1/m_2$ (where $m_1$ is
the small particle and $m_2$ is the black hole). In the test-mass
limit the redshift factor is given by the Schwarzschild value,
\begin{equation}\label{uTschw}
u_\text{Schw}^T = \frac{1}{\sqrt{1-3y}}\,,
\end{equation}
where $y=(G m_2\Omega/c^3)^{2/3}$ is the frequency-related parameter
associated with the motion of the test-mass particle around the black
hole. The self-force part to the redshift factor $u_\text{SF}^T$ is
then defined as $u_1^T =
u_\text{Schw}^T+q\,u_\text{SF}^T+\mathcal{O}(q^2)$. We finally find
that the half-integral conservative contributions therein up to 2PN
relative order are
\begin{equation}\label{uTSF}
u_\text{SF}^T = - y - 2 y^2 -5 y^3 +\cdots -
\frac{13696}{525}\pi\,y^{13/2} + \frac{81077}{3675}\pi\,y^{15/2} +
\frac{82561159}{467775}\pi\,y^{17/2} + \cdots\,,
\end{equation}
where we have written only the relative 2PN terms relevant to our
next-to-next-to-leading order calculation, \textit{i.e.} the
Newtonian, 1PN and 2PN terms for the dominant effects, and the 5.5PN,
6.5PN and 7.5PN terms for the half-integral conservative corrections,
with all the other terms, not computed in the present work, indicated
by ellipsis.\footnote{The sign of the Newtonian term in Eq.~(5.18) of
  Paper~I should be changed and read $-y$.} The result~\eqref{uTSF} is
in full agreement with results derived by gravitational self-force
methods, either numerical, semi-analytical or purely
analytical~\cite{SFW14, BiniD14a, BiniD14b}.

Let us emphasize again that the result~\eqref{uTSF} has been achieved
from the traditional post-Newtonian approach. Contrary to various
analytical and numerical self-force calculations~\cite{SFW14,
  BiniD14a, BiniD14b} the PN approach is completely general,
\textit{i.e.} it is not tuned to a particular type of source as it is
applicable to any extended post-Newtonian source with spatial
compact support. It is remarkable that this general method can
nevertheless be specialized to such degree that it is able to control
terms up to the very high order 7.5PN.

With the post-Newtonian coefficients in the redshift
factor~\eqref{uTSF}, one can straightforwardly obtain, by making use
of the first law of black hole binary mechanics~\cite{LBW12}, the
corresponding coefficients in the PN binding energy and angular
momentum of the system~\cite{LBB12} and the most important
effective-one-body (EOB) potential~\cite{BBL11, Akcay12}.

In the remainder of this paper, we first discuss vacuum solutions in
the exterior zone (Section~\ref{sec:veq-ext}). Then we investigate
tail-of-tail terms in the near zone (Section~\ref{sec:tail2}), listing
the terms which need to be evaluated, and introduce a gauge
transformation to shorten the subsequent calculation. In
Section~\ref{sec:nonlinear}, we set up the PN iteration of
tails of tails, then compute the quadratic and cubic contributions in
turn. We end with a brief discussion of our results
(Section~\ref{sect:disc}), with Appendix~\ref{app:alt} providing an
alternative derivation of some key results, and with
Appendix~\ref{app:source} providing the source terms required for our
tail-of-tail calculations.

\section{Solving the Einstein equations in the exterior zone}
\label{sec:veq-ext}

In the present paper we shall continue and extend the method of
Paper~I. Namely we compute a series of non-linear tail effects in the
exterior vacuum region around a general isolated source. We show that
a crucial piece in the expansion of these non-linear tails can be
extended using a matching argument from the near zone of the source to
the inner region of the source, while the other pieces will not
contribute to the half integral post-Newtonian orders in which we are
interested. This crucial piece is then specialized to the case of
point particle binaries and evaluated at the very location of one of
the particles. Finally the corresponding metric is inserted into the
redshift factor of that particle and the small mass-ratio limit is
computed in order to obtain the self-force prediction which is
meaningfully compared to direct analytical or numerical self-force
calculations.

The vacuum exterior field of a general source is computed using the
multipolar-post-Minkowskian (MPM) algorithm~\cite{BD86, B98quad,
  B98tail}, \textit{i.e.} decomposed into multipolar spherical
harmonics and iterated in a non-linear or post-Minkowskian way. Using
harmonic coordinates, the equation that we have to solve at each
post-Minkowskian order is a (flat) d'Alembertian equation for the
components of the gothic metric deviation, whose right hand-side is
known from previous iterations. Furthermore, if we project out that
equation on a basis of multipolar spherical harmonics with multipole
index $\ell$, we end up with solving a generic equation of the type
\begin{equation}\label{Dalembert}
\Box u_L(\mathbf{x},t) = \hat{n}_L S(r,t-r/c) \,.
\end{equation}
Here $\Box\equiv\eta^{\mu\nu}\partial_\mu\partial_\nu$ is the flat
space-time d'Alembertian operator, $r=\vert\mathbf{x}\vert$ is the
coordinate distance from the field point to the origin located inside
the matter source, and $\hat{n}_L$ is a symmetric-trace-free (STF)
product of $\ell$ unit vectors $n_i=x_i/r$, which is equivalent to the
usual basis of spherical harmonics.\footnote{For STF tensors we use
  the same notation as in Paper~I: $L = i_1 \cdots i_\ell$ denotes a
  multi-index composed of $\ell$ spatial indices ranging from 1 to 3;
  similarly $L-1 = i_1 \cdots i_{\ell-1}$; $\partial_L =
  \partial_{i_1} \cdots \partial_{i_\ell}$ is the product of $\ell$
  partial derivatives $\partial_i \equiv \partial / \partial x^i$;
  $x_L = x_{i_1} \cdots x_{i_\ell}$ is the product of $\ell$ spatial
  positions $x_i$; $n_L = n_{i_1} \cdots n_{i_\ell}$ is the
  product of $\ell$ unit vectors $n_i=x_i/r$; the STF projection is
  indicated with a hat, \textit{i.e.} $\hat{x}_L \equiv
  \text{STF}[x_L]$, $\hat{n}_L \equiv \text{STF}[n_L]$,
  $\hat{\partial}_L \equiv \text{STF}[\partial_L]$ (for instance
  $\hat{\partial}_{ij}=\partial_{ij}-\frac{1}{3}\delta_{ij}\Delta$),
  or sometimes with angular brackets surrounding the indices,
  \textit{e.g.} $x_{\langle i_\ell}\partial_{L-1\rangle}\equiv
  \text{STF}[x_{i_\ell}\partial_{L-1}]$; in the case of summed-up
  multi-indices $L$, we do not write the $\ell$ summations from 1 to 3
  over the dummy indices.} The solution of Eq.~\eqref{Dalembert} for a
source term $S$ which tends to zero sufficiently rapidly when $r\to 0$
(see the precise conditions in Ref.~\cite{BD86}) reads
\begin{equation}\label{uL}
u_L(\mathbf{x},t) = c\int_{-\infty}^{t-r/c}\ud s\,\hat{\partial}_L
\left\{\frac{1}{r}\left[R\Bigl(\frac{t-s-r/c}{2},s\Bigr)
  -R\Bigl(\frac{t-s+r/c}{2},s\Bigr)\right]\right\}\,,
\end{equation}
where $R(\rho, s)$ denotes some intermediate function defined in terms
of the source by
\begin{equation}\label{R}
R\left(\rho,s\right) = \rho^\ell\int_0^\rho\ud
\lambda\,\frac{(\rho-\lambda)^\ell}{\ell!}
\left(\frac{2}{\lambda}\right)^{\ell-1}\!\!S(\lambda,s)\,.
\end{equation}
For the present work, since we shall perform a matching of this
solution to the inner field of a post-Newtonian source, we shall need
the expansion of the solution~\eqref{uL} in the near zone,
\textit{i.e.} when $r\to 0$ formally. Denoting with an overbar the
formal expansion when $r\to 0$ we can write the following crucial
formula~\cite{B93}:
\begin{equation}\label{formuleZP}
\overline{u_L(\mathbf{x},t)} =
\hat{\partial}_L\left\{\overline{\frac{G(t-r/c)-G(t+r/c)}{r}}\right\}
+ \Box^{-1}_\text{inst}\bigl[\hat{n}_L \overline{S(r,t-r/c)}\bigr]\,.
\end{equation}
The second term in that formula represents a particular solution of
the equation~\eqref{Dalembert}, in the form of an expansion when $r\to
0$, and given by the so-called operator of the instantaneous
potentials defined by
\begin{equation}\label{Iinst}
\Box^{-1}_\text{inst}\bigl[\hat{n}_L \overline{S(r,t-r/c)}\bigr] =
\sum_{i=0}^{+\infty}\left(\frac{\partial}{c\partial
  t}\right)^{2i}\Delta^{-1-i}\bigl[\hat{n}_L
  \overline{S(r,t-r/c)}\bigr]\,.
\end{equation}
Note that such operator acts directly (term by term) on the formal
expansion of the source when $r\to 0$, given by the usual Tayor
expansion of the retardation $t-r/c$, and does not integrate over time
(hence the adjective ``instantaneous''); see Ref.~\cite{B93} for the
proof and more details about the iterated Poisson operator in
Eq.~\eqref{Iinst}.

The point, proved in the Appendix of Paper~I, is that the second term
in Eq.~\eqref{formuleZP} always contributes to \textit{integral}
post-Newtonian approximations, and thus can be safely ignored when
looking at the half-integral approximations. We shall check in
Appendix~\ref{app:source} below that the proof of Paper~I is still
applicable to the extended calculation performed here. Thus all the
effects we are looking for come from the first term in
Eq.~\eqref{formuleZP}, which is a homogeneous solution of the wave
equation of the type retarded minus advanced and is parametrized by
the function
\begin{equation}\label{G}
G\left(u\right) = c\int_{-\infty}^u \ud
s\,R\left(\frac{u-s}{2},s\right)\,.
\end{equation}
Note that the retarded-minus-advanced solution is regular when $r\to
0$ and can therefore be directly extended by matching inside the
source. The purpose is to compute the function $G$ given the
generic form of the source term $S$ we need. As in Paper~I we apply
Eq.~\eqref{G}, together with Eq.~\eqref{R}, to source terms made of
the requisite tails, that is, non-local in time (hereditary) terms
having the form
\begin{equation}\label{source}
S(r,t-r/c) = r^{B-k}\int_1^{+\infty}\ud x\,Q_m(x)F(t-r x/c)\,,
\end{equation}
where $F$ denotes some time derivative of a multipole moment, $k$ and
$m$ are integers, and $Q_m(x)$ is the Legendre function of the second
kind, with branch cut from $-\infty$ to $1$, explicitly given in
terms of the usual Legendre polynomial $P_m(x)$ by
\begin{equation}\label{Qm}
Q_m(x) = \frac{1}{2} P_m (x) \,\ln \left(\frac{x+1}{x-1} \right)-
\sum^{m}_{ j=1} \frac{1}{j} P_{m-j}(x) P_{j-1}(x)\,.
\end{equation}

Besides the hereditary source terms~\eqref{source} we need also to
include the case of instantaneous (non-tail) terms of the type
$S(r,t-r/c) = r^{B-k}\,F(t-r/c)$, but these are immediately deduced from
the hereditary case~\eqref{source} by replacing formally $Q_m(x)$ by
the truncated delta-function defined by
$\delta_+(x-1)=Y(x-1)\delta(x-1)$, where $Y$ and $\delta$ are the
usual Heaviside and delta functions. Hence we can handle all the terms
given for completeness in Appendix~\ref{app:source}.

Note that we systematically include inside the source
term~\eqref{source} a regularization factor $r^B$, where $B$ is a
complex parameter destined to tend to zero at the end of the
calculation. The presence of this factor ensures, when the real part
$\Re(B)$ is large enough, that the source term tends sufficiently
rapidly toward zero when $r\to 0$, so the applicability conditions of
the integration formulas~\eqref{uL} and~\eqref{formuleZP} are
fulfilled (see Refs.~\cite{BD86, B93}). From the initial domain of the
complex plane where $\Re(B)$ is large enough, we extend the validity
of the formulas by analytic continuation to any complex $B$-values
except isolated poles at integer values of $B$.

Plugging the source term~\eqref{source} into Eq.~\eqref{R}, and then
substituting~\eqref{R} into Eq.~\eqref{G}, we obtained in Paper~I a
more tractable expression of the function $G$ that parametrizes the
term of interest to us in Eq.~\eqref{formuleZP}, namely
\begin{equation}\label{Gres}
G\left(u\right) = c^{B+\ell-k+3}C_{k,\ell,m}(B)\int_0^{+\infty}\ud
\tau\,\tau^B F^{(k-\ell-2)}(u-\tau)\,.
\end{equation}
Always implicit in expressions such as Eq.~\eqref{Gres} is that we
perform the Laurent expansion of the result when $B\to 0$ and then
pick up the finite part of that expansion, \textit{i.e.} the
coefficient of the zero-th power of $B$. Depending on the relative
values of $k$ and $\ell$ (namely the power of $1/r$ and the multipole
order of the term in question), the function $F$ in Eq.~\eqref{Gres}
will appear either multi time-differentiated or multi time-integrated,
which we indicate in both cases by the superscript $(p)$ where
$p=k-\ell-2$ can be positive or negative; the formula~\eqref{Gres} is
valid in either cases. The $B$-dependent coefficient $C_{k,\ell,m}$ in
Eq.~\eqref{Gres} reads
\begin{equation}\label{CB}
C_{k,\ell,m}(B) =
\frac{2^\ell}{\ell!}\frac{\Gamma(B-k+\ell+3)}{\Gamma(B+1)}
\int_0^{+\infty}\ud y\,Q_m(1+y)\int_0^1\ud
z\,\frac{z^{B-k-\ell+1}(1-z)^\ell}{(2+y z)^{B-k+\ell+3}}\,,
\end{equation}
where $\Gamma$ is the usual Eulerian function; see Paper~I for more
details. An alternative form of Eq.~\eqref{CB}, also derived in
Paper~I, is
\begin{equation}\label{CBalt}
C_{k,\ell,m}(B) = \frac{\Gamma(B-k-\ell+2)}{2\Gamma(B+1)}
\sum_{i=0}^\ell\frac{(\ell+i)!}{i!(\ell-i)!}
\frac{\Gamma(B-k+\ell+3)}{\Gamma(B-k+i+3)} \int_0^{+\infty}\ud
y\,\left(\frac{y}{2}\right)^i\frac{Q_m(1+y)}{(2+y)^{B-k+2}}\,.
\end{equation}

In order to control the tails present in the function $G$, and which
are responsible for the half-integral post-Newtonian approximations
(Paper~I), we need to control the pole parts when $B\to 0$ of the
expressions~\eqref{CB} or~\eqref{CBalt}, see Eqs.~(4.10)--(4.12) in
Paper~I. In particular, when only simple poles $\propto 1/B$ appear
which will always be the case in the present paper, the tail part of
the function $G$ is given by
\begin{equation}\label{Gtailpole}
G^\text{tail}(u) =
c^{\ell-k+3}\alpha_{k,\ell,m}^{(-1)}\int_0^{+\infty}\ud
\tau\,\ln\tau\,F^{(k-\ell-2)}(u-\tau)\,,
\end{equation}
where $\alpha_{k,\ell,m}^{(-1)}$ denotes the residue (\textit{i.e.}
coefficient of $1/B$) in the Laurent expansion of the coefficient
$C_{k,\ell,m}(B)$ when $B\to 0$. The residue can be obtained either by
carefully expanding Eqs.~\eqref{CB} or~\eqref{CBalt} when $B\to 0$ as
was done in Paper~I, or by using a powerful alternative method,
described in Appendix~\ref{app:alt}, which is especially tuned to pick
up directly and rapidly the required pole parts.

The tail integral~\eqref{Gtailpole} involves as usual a logarithmic
kernel. Note that we keep the argument of the logarithm without a
constant to adimensionalize it, \textit{e.g.} $\ln(\tau/P)$, because
any constant $P$ will yield an intantaneous (non-tail) term that is
safely ignored here.

\section{Tails of tails in the near-zone}
\label{sec:tail2}

\subsection{Expressions in harmonic coordinates}
\label{sec:harm}

A straightforward extension of the analysis of Paper~I (see Sec.~II
there) shows that in order to control the half-integral post-Newtonian
coefficients up to next-to-next-to-leading order, namely 2PN beyond
the leading-order 5.5PN coefficient obtained in Paper~I, we need to
compute the tails of tails associated with the \textit{mass-type}
quadrupole, octupole and hexadecapole moments, and with the
\textit{current-type} quadrupole and octupole moments. In the notation
of Paper~I, this means that we have to take into account the multipole
interactions $M \times M \times I_{ij}$ (that one was sufficient for
Paper~I), $M \times M \times I_{ijk}$ and $M \times M \times I_{ijkl}$
for mass moments, as well as $M \times M \times J_{ij}$ and $M \times M
\times J_{ijk}$ for current moments. As will be discussed in
Sec.~\ref{sec:nonlinear}, those interactions represent only the
``seeds'' for a subsequent post-Newtonian iteration, formally
involving higher non-linear multipole interactions.

For all the ``seed'' multipole interactions we only need the functions
$G$ parametrizing the regular retarded-minus-advanced homogeneous
solutions in Eq.~\eqref{formuleZP}. They are obtained from applying
Eqs.~\eqref{Gres}--\eqref{CBalt} to each one of the source terms
corresponding to these multipole interactions. The computation is
straightforward, and for completeness we present in
Appendix~\ref{app:source} the complete expressions of the required
source terms, extending Eqs.~(3.4)--(3.5) of Paper~I. Typically all
the coefficients in Eqs.~\eqref{sourceinstI2}--\eqref{sourcetailJ3} of
Appendix~\ref{app:source} contribute to the final results. To ease the
notation we use the following shorthand for an elementary monopolar
retarded-minus-advanced homogeneous wave,
\begin{equation}\label{def}
\bigl\{G(t)\bigr\} \equiv \frac{G(t-r/c) - G(t+r/c)}{r} \,.
\end{equation}
Corresponding multipolar retarded-minus-advanced waves are obtained by
applying STF partial space multi-derivative operators
$\hat{\partial}_L$ (with multipolarity $\ell$). The near-zone
expansion when $r\to 0$ of such multipolar waves is given by the Taylor
expansion as
\begin{equation}\label{defZP}
\overline{\hat{\partial}_L\bigl\{G(t)\bigr\}} =
-2\sum_{k=0}^{+\infty}\frac{\hat{x}_L\,r^{2k}}{(2k)!!(2k+2\ell+1)!!}
\frac{G^{(2k+2\ell+1)}(t)}{c^{2k+2\ell+1}}\,.
\end{equation}

Extending Eqs.~(5.1) of Paper~I, we present the multipolar
tail-of-tail interactions corresponding to the first term of
Eq.~\eqref{formuleZP}, for each of the components of the gothic metric
deviation $h^{\mu\nu}\equiv\sqrt{-g}g^{\mu\nu}-\eta^{\mu\nu}$ in
harmonic gauge, such that $\partial_\nu h^{\mu\nu}=0$. All these
contributions are built from the source terms given in
Eqs.~\eqref{sourceinstI2}--\eqref{sourcetailJ3}.
\begin{itemize}
\item Mass quadrupole moment:
\begin{subequations}\label{hI2}
\begin{align}
  (h^{00})_{M \times M \times I_{ij}} &=
  \frac{116}{21}\frac{G^3M^2}{c^{8}}\int_0^{+\infty}\ud\,\tau\ln\tau
  \,\partial_{ab}\bigl\{I_{ab}^{(3)}(t-\tau)\bigr\} \,,\\
  (h^{0i})_{M \times M \times I_{ij}} &=
  \frac{4}{105}\frac{G^3M^2}{c^{7}}\int_0^{+\infty}\ud\,\tau\ln\tau
  \,\hat{\partial}_{iab}\bigl\{I_{ab}^{(2)}(t-\tau) \bigr\}\nonumber\\&
  -\frac{416}{75}\frac{G^3M^2}{c^{9}}\int_0^{+\infty}\ud\,\tau\ln\tau
  \,\partial_{a}\bigl\{I_{ia}^{(4)}(t-\tau)\bigr\}\,,\\
(h^{ij})_{M
    \times M \times I_{ij}} &=
  -\frac{32}{21}\frac{G^3M^2}{c^{8}}\int_0^{+\infty}\ud\,\tau\ln\tau
  \,\delta_{ij}\partial_{ab}\bigl\{I_{ab}^{(3)}(t-\tau)\bigr\}\nonumber\\&
  +\frac{104}{35}\frac{G^3M^2}{c^{8}}\int_0^{+\infty}\ud\,\tau\ln\tau
  \,\hat{\partial}_{a(i}\bigl\{I_{j)a}^{(3)}(t-\tau) \bigr\}\nonumber\\&
  +\frac{76}{15}\frac{G^3M^2}{c^{10}}\int_0^{+\infty}\ud\,\tau\ln\tau
  \,\bigl\{I_{ij}^{(5)}(t-\tau)\bigr\}\,.
\end{align}
\end{subequations}
\item Mass octupole:
\begin{subequations}\label{hI3}
\begin{align}
  (h^{00})_{M \times M \times I_{ijk}} &= -
  \frac{328}{315}\frac{G^3M^2}{c^{8}}\int_0^{+\infty}\ud\,\tau\ln\tau
  \,\partial_{abc}\bigl\{I_{abc}^{(3)}(t-\tau)\bigr\}\,,\\
  (h^{0i})_{M \times M \times I_{ijk}} &=
  -\frac{2}{315}\frac{G^3M^2}{c^{7}}\int_0^{+\infty}\ud\,\tau\ln\tau
  \,\hat{\partial}_{iabc}\bigl\{I_{abc}^{(2)}(t-\tau)\bigr\}\nonumber\\&
  +\frac{256}{245}\frac{G^3M^2}{c^{9}}\int_0^{+\infty}\ud\,\tau\ln\tau
  \,\partial_{ab}\bigl\{I_{iab}^{(4)}(t-\tau)\bigr\}\,,
\\(h^{ij})_{M
    \times M \times I_{ijk}} &=
  \frac{8}{35}\frac{G^3M^2}{c^{8}}\int_0^{+\infty}\ud\,\tau\ln\tau
  \,\delta_{ij}\partial_{abc}\bigl\{I_{abc}^{(3)}(t-\tau)\bigr\}\nonumber\\&
  -\frac{4}{9}\frac{G^3M^2}{c^{8}}\int_0^{+\infty}\ud\,\tau\ln\tau
  \,\hat{\partial}_{ab(i}\bigl\{I_{j)ab}^{(3)}(t-\tau)\bigr\}\nonumber\\&
  -\frac{316}{315}\frac{G^3M^2}{c^{10}}\int_0^{+\infty}\ud\,\tau\ln\tau
  \,\partial_{a}\bigl\{I_{ija}^{(5)}(t-\tau)\bigr\}\,.
\end{align}
\end{subequations}
\item Mass hexadecapole:
\begin{subequations}\label{hI4}
\begin{align}
  (h^{00})_{M \times M \times I_{ijkl}} &=
  \frac{1898}{10395}\frac{G^3M^2}{c^{8}}\int_0^{+\infty}\ud\,\tau\ln\tau
  \,\partial_{abcd}\bigl\{I_{abcd}^{(3)}(t-\tau)\bigr\}\,,\\ (h^{0i})_{M
    \times M \times I_{ijkl}} &=
  \frac{1}{1155}\frac{G^3M^2}{c^{7}}\int_0^{+\infty}\ud\,\tau\ln\tau
  \,\hat{\partial}_{iabcd}\bigl\{I_{abcd}^{(2)}(t-\tau)\bigr\}\nonumber\\&
  -\frac{173}{945}\frac{G^3M^2}{c^{9}}\int_0^{+\infty}\ud\,\tau\ln\tau
  \,\partial_{abc}\bigl\{I_{iabc}^{(4)}(t-\tau)\bigr\}\,,\\(h^{ij})_{M
    \times M \times I_{ijkl}} &= -
  \frac{23}{693}\frac{G^3M^2}{c^{8}}\int_0^{+\infty}\ud\,\tau\ln\tau
  \,\delta_{ij}\partial_{abcd}\bigl\{I_{abcd}^{(3)}(t-\tau)\bigr\}\nonumber\\&
  +\frac{32}{495}\frac{G^3M^2}{c^{8}}\int_0^{+\infty}\ud\,\tau\ln\tau
  \,\hat{\partial}_{abc(i}\bigl\{I_{j)abc}^{(3)}(t-\tau)\bigr\}\nonumber\\&
  +\frac{169}{945}\frac{G^3M^2}{c^{10}}\int_0^{+\infty}\ud\,\tau\ln\tau
  \,\partial_{ab}\bigl\{I_{ijab}^{(5)}(t-\tau)\bigr\}\,.
\end{align}
\end{subequations}
\item Current quadrupole:\footnote{Underlined indices mean that they
  should be excluded from the symmetrization
  $T_{(ij)}=\frac{1}{2}(T_{ij}+T_{ji})$.}
\begin{subequations}\label{hJ2}
\begin{align}
  (h^{00})_{M \times M \times J_{ij}} &= 0\,,\\
  (h^{0i})_{M \times M \times J_{ij}} &=
  \frac{296}{105}\frac{G^3M^2}{c^{9}}
\int_0^{+\infty}\ud\,\tau\ln\tau
  \, \varepsilon_{iab}\,\partial_{bc}\bigl\{J_{ac}^{(3)}(t-\tau)\bigr\} \,,\\
(h^{ij})_{M\times M \times J_{ij}} &=
  -\frac{64}{315}\frac{G^3M^2}{c^{8}} \int_0^{+\infty}\ud\,\tau\ln\tau
  \,\varepsilon_{ab(i}\,\hat{\partial}_{j)bc}
  \bigl\{J_{ac}^{(2)}(t-\tau)\bigr\}\nonumber\\&
  -\frac{1232}{225}\frac{G^3M^2}{c^{10}}
  \int_0^{+\infty}\ud\,\tau\ln\tau \,
  \varepsilon_{ab(i}\,\partial_{\underline{b}}
  \bigl\{J_{j)a}^{(4)}(t-\tau)\bigr\}\,.
\end{align}
\end{subequations}
\item Current octupole:
\begin{subequations}\label{hJ3}
\begin{align}
  (h^{00})_{M \times M \times J_{ijk}} &= 0\,,\\ (h^{0i})_{M \times M
    \times J_{ijk}} &= - \frac{68}{105}\frac{G^3M^2}{c^{9}}
  \int_0^{+\infty}\ud\,\tau\ln\tau
  \,\varepsilon_{iab}\,\partial_{bcd}\bigl\{J_{acd}^{(3)}(t-\tau)\bigr\}
  \,,\\ (h^{ij})_{M\times M \times J_{ijk}} &=
  \frac{2}{35}\frac{G^3M^2}{c^{8}} \int_0^{+\infty}\ud\,\tau\ln\tau
  \,\varepsilon_{ab(i}\,\hat{\partial}_{j)bcd}
  \bigl\{J_{acd}^{(2)}(t-\tau)\bigr\}\nonumber\\&
  +\frac{922}{735}\frac{G^3M^2}{c^{10}}
  \int_0^{+\infty}\ud\,\tau\ln\tau
  \,\varepsilon_{ab(i}\,\partial_{\underline{bc}}
  \bigl\{J_{j)ac}^{(4)}(t-\tau)\bigr\}\,.
\end{align}
\end{subequations}
\end{itemize}

\subsection{Application of a gauge transformation}
\label{sec:changegauge}

As noticed in Paper~I the tail-of-tail term $M \times M \times I_{ij}$
given by Eq.~\eqref{hI2} is to be iterated at higher non-linear order
as there are some post-Newtonian terms which contribute at the same
level coming from higher non-linear iterations. However it was found
that the details of that non-linear iteration depend on the adopted
coordinate system. In Paper~I two computations of the 5.5PN
coefficient were made. One in the standard harmonic coordinate system,
based on the previous expressions~\eqref{hI2}, and one in an
alternative coordinate system in which the 5.5PN terms in the $0i$ and
$ij$ components of the metric are ``transferred'' to the $00$
component at that order. This alternative coordinate system has the
great advantage that it considerably simplifies the subsequent
non-linear iteration. Actually, it was found in Paper~I that at 5.5PN
order in this coordinate system, there is no need to perform the
non-linear iteration. Such a coordinate system is analogous to the Burke
\& Thorne coordinate system~\cite{BuTh70, Bu71} (see also~\cite{MTW}),
in which the complete radiation reaction force at the 2.5PN order is
linear, with non-linear contributions arising only at higher
post-Newtonian orders.

In the present paper we shall systematically work in the alternative
non-harmonic coordinate system so designed that it minimizes (but, at
such high 7.5PN order, does not suppress) the need for controlling
non-linear contributions. Even in that optimized gauge we shall find
that the non-linear contributions are numerous and require two
iterations. We did not attempt to perform these non-linear iterations
in harmonic coordinates. Since the redshift factor we compute
\textit{in fine} is gauge invariant we are allowed to use whatever
coordinate system we like. Thus we proceed with introducing
appropriate gauge transformation vectors $\eta^\mu$ to be applied to
each of the multipolar pieces presented in Sec.~\ref{sec:harm}. The
complete gauge transformation is of course the sum of each of the
separate multipolar pieces. At leading 5.5PN order the mass quadrupole
piece agrees with Eqs.~(5.11) of Paper~I, except that here we do not
yet focus our attention on the conservative part of the dynamics; a
split between conservative and dissipative parts will be made at a
later stage, see Eqs.~\eqref{Fijcons}. Note also that the above gauge
vectors generalize those of Paper~I not only because they involve more
multipole interactions but also because they include all
post-Newtonian terms, \textit{i.e.} complete series expansions such as
Eq.~\eqref{defZP}.
\begin{itemize}
\item Mass quadrupole:
\begin{subequations}\label{etaI2}
\begin{align}
  (\eta^{0})_{M \times M \times I_{ij}} &=
  \frac{77}{15}\frac{G^3M^2}{c^{7}}\int_0^{+\infty}\ud\,\tau\ln\tau
  \,\partial_{ab}\bigl\{I_{ab}^{(2)}(t-\tau)\bigr\}\,,\\
  (\eta^{i})_{M \times M \times I_{ij}} &=
  -\frac{107}{3}\frac{G^3M^2}{c^{6}}\int_0^{+\infty}\ud\,\tau\ln\tau
  \,\hat{\partial}_{iab}\bigl\{I_{ab}^{(1)}(t-\tau)\bigr\}\nonumber\\&
  +\frac{38}{5}\frac{G^3M^2}{c^{8}}\int_0^{+\infty}\ud\,\tau\ln\tau
  \,\partial_{a}\bigl\{I_{ia}^{(3)}(t-\tau)\bigr\}\,.
\end{align}
\end{subequations}
\item Mass octupole:
\begin{subequations}\label{etaI3}
\begin{align}
  (\eta^{0})_{M \times M \times I_{ijk}} &=
 - \frac{461}{945}\frac{G^3M^2}{c^{7}}\int_0^{+\infty}\ud\,\tau\ln\tau
  \,\partial_{abc}\bigl\{I_{abc}^{(2)}(t-\tau)\bigr\}\,,\\
  (\eta^{i})_{M \times M \times I_{ijk}} &=
  \frac{13}{3}\frac{G^3M^2}{c^{6}}\int_0^{+\infty}\ud\,\tau\ln\tau
  \,\hat{\partial}_{iabc}\bigl\{I_{abc}^{(1)}(t-\tau)\bigr\}\nonumber\\&
  -\frac{79}{63}\frac{G^3M^2}{c^{8}}\int_0^{+\infty}\ud\,\tau\ln\tau
  \,\partial_{ab}\bigl\{I_{iab}^{(3)}(t-\tau)\bigr\}\,.
\end{align}
\end{subequations}
\item Mass hexadecapole:
\begin{subequations}\label{etaI4}
\begin{align}
  (\eta^{0})_{M \times M \times I_{ijkl}} &=
 \frac{29}{504}\frac{G^3M^2}{c^{7}}\int_0^{+\infty}\ud\,\tau\ln\tau
  \,\partial_{abcd}\bigl\{I_{abcd}^{(2)}(t-\tau)\bigr\}\,,\\
  (\eta^{i})_{M \times M \times I_{ijkl}} &=
  - \frac{1571}{2520}\frac{G^3M^2}{c^{6}}\int_0^{+\infty}\ud\,\tau\ln\tau
  \,\hat{\partial}_{iabcd}\bigl\{I_{abcd}^{(1)}(t-\tau)\bigr\}\nonumber\\&
  +\frac{169}{810}\frac{G^3M^2}{c^{8}}\int_0^{+\infty}\ud\,\tau\ln\tau
  \,\partial_{abc}\bigl\{I_{iabc}^{(3)}(t-\tau)\bigr\}\,.
\end{align}
\end{subequations}
\item Current quadrupole:
\begin{subequations}\label{etaJ2}
\begin{align}
  (\eta^{0})_{M \times M \times J_{ij}} &= 0\,,\\
  (\eta^{i})_{M \times M \times J_{ij}} &=
  - \frac{616}{45}\frac{G^3M^2}{c^{8}}
\int_0^{+\infty}\ud\,\tau\ln\tau
  \, \varepsilon_{iab} \partial_{bc}\bigl\{J_{ac}^{(2)}(t-\tau)\bigr\} \,.
\end{align}
\end{subequations}
\item Current octupole:
\begin{subequations}\label{etaJ3}
\begin{align}
  (\eta^{0})_{M \times M \times J_{ijk}} &= 0\,,\\
  (\eta^{i})_{M \times M \times J_{ijk}} &=
  \frac{461}{210}\frac{G^3M^2}{c^{8}} \int_0^{+\infty}\ud\,\tau\ln\tau
  \, \varepsilon_{iab}
  \partial_{bcd}\bigl\{J_{acd}^{(2)}(t-\tau)\bigr\} \,.
\end{align}
\end{subequations}
\end{itemize}

Applying the latter linear gauge transformations we obtain new
expressions for the gothic metric coefficients, say
${h'}^{\mu\nu}$. Our convention is that (for each multipole component)
\begin{equation}
{h'}^{\mu\nu}=h^{\mu\nu}-\partial^\mu\eta^\nu-\partial^\nu\eta^\mu +
\eta^{\mu\nu}\partial_\rho\eta^\rho\,.
\end{equation}
The nice property of the metric in the new gauge is that the number
$\ell$ of STF spatial derivatives $\hat{\partial}_L$ for each
multipole is maximal, and equal to $\ell=m+s$ for mass moments and
$\ell=m+s-1$ for current moments, where $m$ is the multipolarity of
the multipole moment in question (\textit{i.e.} $I_M$ or $J_M$) and
$s$ is the number of spatial indices in the gothic metric
(\textit{i.e.} $s=0,1,2$ according to whether $\mu\nu=00, 0i, ij$). From
Eq.~\eqref{defZP} we see that maximizing the number of STF derivatives
means pushing to the maximum the leading PN order, and therefore
minimizing the need of non-linear iterations at a given PN level.

\begin{itemize}
\item Mass quadrupole:
\begin{subequations}\label{hI2p}
\begin{align}
  ({h'}^{00})_{M \times M \times I_{ij}} &=
  \frac{856}{35}\frac{G^3M^2}{c^{8}}\int_0^{+\infty}\ud\,\tau\ln\tau
  \,\partial_{ab}\bigl\{I_{ab}^{(3)}(t-\tau)\bigr\}\,,\\
  ({h'}^{0i})_{M \times M \times I_{ij}} &=
  -\frac{856}{21}\frac{G^3M^2}{c^{7}}\int_0^{+\infty}\ud\,\tau\ln\tau
  \,\hat{\partial}_{iab}\bigl\{I_{ab}^{(2)}(t-\tau)\bigr\}\,,\\
({h'}^{ij})_{M
    \times M \times I_{ij}} &=
  \frac{214}{3}\frac{G^3M^2}{c^{6}}\int_0^{+\infty}\ud\,\tau\ln\tau
  \,\hat{\partial}_{ijab}\bigl\{I_{ab}^{(1)}(t-\tau)\bigr\}\,.
\end{align}
\end{subequations}
\item Mass octupole:
\begin{subequations}\label{hI3p}
\begin{align}
  ({h'}^{00})_{M \times M \times I_{ijk}} &=
 - \frac{520}{189}\frac{G^3M^2}{c^{8}}\int_0^{+\infty}\ud\,\tau\ln\tau
  \,\partial_{abc}\bigl\{I_{abc}^{(3)}(t-\tau)\bigr\}\,,\\
  ({h'}^{0i})_{M \times M \times I_{ijk}} &=
  \frac{130}{27}\frac{G^3M^2}{c^{7}}\int_0^{+\infty}\ud\,\tau\ln\tau
  \,\hat{\partial}_{iabc}\bigl\{I_{abc}^{(2)}(t-\tau)\bigr\}\,,\\
({h'}^{ij})_{M \times M \times I_{ijk}} &=
  -\frac{26}{3}\frac{G^3M^2}{c^{6}}\int_0^{+\infty}\ud\,\tau\ln\tau
  \,\hat{\partial}_{ijabc}\bigl\{I_{abc}^{(1)}(t-\tau)\bigr\}\,.
\end{align}
\end{subequations}
\item Mass hexadecapole:
\begin{subequations}\label{hI4p}
\begin{align}
  ({h'}^{00})_{M \times M \times I_{ijkl}} &=
 \frac{1571}{4158}\frac{G^3M^2}{c^{8}}\int_0^{+\infty}\ud\,\tau\ln\tau
  \,\partial_{abcd}\bigl\{I_{abcd}^{(3)}(t-\tau)\bigr\}\,,\\
  ({h'}^{0i})_{M \times M \times I_{ijkl}} &=
  - \frac{1571}{2310}\frac{G^3M^2}{c^{7}}\int_0^{+\infty}\ud\,\tau\ln\tau
  \,\hat{\partial}_{iabcd}\bigl\{I_{abcd}^{(2)}(t-\tau)\bigr\}\,,\\
({h'}^{ij})_{M \times M \times I_{ijkl}} &=
  \frac{1571}{1260}\frac{G^3M^2}{c^{6}}\int_0^{+\infty}\ud\,\tau\ln\tau
  \,\hat{\partial}_{ijabcd}\bigl\{I_{abcd}^{(1)}(t-\tau)\bigr\}\,.
\end{align}
\end{subequations}
\item Current quadrupole:
\begin{subequations}\label{hJ2p}
\begin{align}
  ({h'}^{00})_{M \times M \times J_{ij}} &= 0\,,\\
  ({h'}^{0i})_{M \times M \times J_{ij}} &=
  -\frac{3424}{315}\frac{G^3M^2}{c^{9}}
  \int_0^{+\infty}\ud\,\tau\ln\tau \, \varepsilon_{iab}
  \partial_{bc}\bigl\{J_{ac}^{(3)}(t-\tau)\bigr\} \,,\\
({h'}^{ij})_{M\times M \times J_{ij}} &=
  \frac{1712}{63}\frac{G^3M^2}{c^{8}} \int_0^{+\infty}\ud\,\tau\ln\tau
  \,\varepsilon_{ab(i}\hat{\partial}_{j)bc}\bigl\{J_{ac}^{(2)}(t-\tau)\bigr\}\,.
\end{align}
\end{subequations}
\item Current octupole:
\begin{subequations}\label{hJ3p}
\begin{align}
  ({h'}^{00})_{M \times M \times J_{ijk}} &= 0\,,\\
  ({h'}^{0i})_{M \times M \times J_{ijk}} &=
  \frac{65}{42}\frac{G^3M^2}{c^{9}}
  \int_0^{+\infty}\ud\,\tau\ln\tau \, \varepsilon_{iab}
  \partial_{bcd}\bigl\{J_{acd}^{(3)}(t-\tau)\bigr\} \,,\\
({h'}^{ij})_{M\times M \times J_{ijk}} &= -
  \frac{13}{3}\frac{G^3M^2}{c^{8}} \int_0^{+\infty}\ud\,\tau\ln\tau
  \,\varepsilon_{ab(i}\hat{\partial}_{j)bcd}
  \bigl\{J_{acd}^{(2)}(t-\tau)\bigr\}\,.
\end{align}
\end{subequations}
\end{itemize}
Notice that ${h'}_{(1)}^{ii}=0$ for all these pieces, which is a nice
feature of the new gauge, shared in fact with the harmonic
gauge. Recall that expressions~\eqref{hI2p}--\eqref{hJ3p} are
regular inside the source and will be valid as they stand at the
location of the particles in a binary system.

\section{Post-Newtonian iteration of tails of tails}
\label{sec:nonlinear}

\subsection{Setting up the iteration}
\label{sec:iter}

As mentioned above, we found in Paper~I that in harmonic coordinates the
computation of the 5.5PN coefficient requires the control of one
non-linear PN iteration, but that no non-linear iteration is needed in
the alternative non-harmonic gauge. To extend the result up to 7.5PN
order, our rationale here is to systematically use the simpler
non-harmonic gauge in which the metric components are given by
Eqs.~\eqref{hI2p}--\eqref{hJ3p}.

In the iteration process we shall have to couple the tail-of-tail
pieces~\eqref{hI2p}--\eqref{hJ3p} with the lower order 1PN
metric. Since the choice of non-harmonic gauge we have made above
affects only the higher order tail-of-tail parts of the metric, we can
take for the 1PN metric the standard form in harmonic coordinates,
given by
\begin{subequations}\label{h1PN}
\begin{align}
h^{00} =& - \frac{4}{c^2}V - \frac{2}{c^4}\left(\hat{W}+4V^2\right) +
\calO\left(\frac{1}{c^{6}}\right)\,,\label{h1PN00}\\
h^{0i} =& - \frac{4}{c^3} V_i +
\calO\left(\frac{1}{c^{5}}\right)\,,\\
h^{ij} =& -
\frac{4}{c^4}\left(\hat{W}_{ij}-\frac{1}{2}\delta_{ij}\hat{W}\right)+
\calO\left(\frac{1}{c^{6}}\right)\,,
\end{align}
\end{subequations}
where we follow our usual notation for appropriate metric potentials
$V$, $V_i$, $\hat{W}_{ij}$ and $\hat{W}=\hat{W}_{kk}$, defined in
Sec.~5.3 of Ref.~\cite{Bliving14} for general post-Newtonian
sources. We then specialize these potentials to point-particle binary
sources. Denoting the masses by $m_A$ ($A=1,2$), the trajectories and
velocities by $y_A^i(t)$ and $v_A^i(t)=\ud y_A^i(t)/\ud t$, the
distances to the field point by
$r_A=\vert\mathbf{x}-\mathbf{y}_A\vert$, and the separation by
$r_{12}=\vert\mathbf{y}_1-\mathbf{y}_2\vert$, we have
\begin{subequations}\label{pot1PN}
\begin{align}
V &= U + \frac{1}{c^2}\,\partial^2_t U_2 +
\calO\left(\frac{1}{c^{3}}\right)\,,\label{V}\\
V_i &= U_i + \calO\left(\frac{1}{c^{2}}\right)\,,\\
\hat{W}_{ij} &= U_{ij} - \delta_{ij} U_{kk} -
\frac{G^2m_1^2}{8}\Bigl[\partial_{ij}\ln r_1 +
  \frac{\delta^{ij}}{r_1^2}\Bigr] -
\frac{G^2m_2^2}{8}\Bigl[\partial_{ij}\ln r_2 +
  \frac{\delta^{ij}}{r_2^2}\Bigr] \nonumber\\ &\qquad\quad - 2 G^2 m_1
m_2 \frac{\partial^2g}{\partial y_1^{(i}\partial y_2^{j)}} +
\calO\left(\frac{1}{c}\right)\,.\label{Wij}
\end{align}
\end{subequations}
Here $U$, $U_i$ and $U_{ij}$ refer to the compact-support parts of the
potentials that are given (consistently with the approximation)
explicitly by
\begin{subequations}\label{potcompact}
\begin{align}
U &= \frac{G \tilde{\mu}_1}{r_1} + \frac{G \tilde{\mu}_2}{r_2}
\,,\label{U}\\ U_2 &= \frac{G \tilde{\mu}_1}{2}\,r_1 + \frac{G
  \tilde{\mu}_2}{2}\,r_2 \,,\label{U2}\\ U_i &= \frac{G
  m_1}{r_1}\,v_1^i + \frac{G m_2}{r_2}\,v_2^i \,,\\ U_{ij} &= \frac{G
  m_1}{r_1}\,v_1^iv_1^j + \frac{G m_2}{r_2}\,v_2^iv_2^j\,.
\end{align}
\end{subequations}
The potential $U$ is 1PN-accurate and we have introduced the effective
time-dependent masses at 1PN order (which are pure functions of time),
\begin{equation}\label{mutilde}
\tilde{\mu}_1 = m_1 \left[1-\frac{G
    m_2}{c^2r_{12}}+\frac{3}{2}\frac{v_1^2}{c^2}\right] \,,
\end{equation}
and $\tilde{\mu}_2$ obtained by exchanging the particle labels. Note
that the potential $U_2$ so defined is the ``super-potential'' of $U$,
in the sense that
\begin{equation}\label{DeltaU2}
\Delta U_2 = U \,.
\end{equation}
Later, we shall systematically make use of the notion of high-order
super-potentials. Finally the non-linear interaction term
in~\eqref{Wij} is expressed by means of the well-known
function~\cite{Fock}
\begin{equation}\label{g}
g=\ln(r_1+r_2+r_{12})\,,
\end{equation}
which is the super-potential of $1/(r_1 r_2)$, \textit{i.e.} $\Delta g
= \frac{1}{r_1 r_2}$ in the sense of distributions. Later we shall
introduce the super-potential of $g$ itself. In Eq.~\eqref{Wij} the
function $g$ is differentiated with respect to the two source points
$y_A^i$ as indicated.

The most important problem we face is the mass quadrupole case, which
we shall need to iterate two times. We need to control the covariant
metric components $g_{00}$, $g_{0i}$ and $g_{ij}$ up to order 7.5PN,
which means $c^{-17}$, $c^{-16}$ and $c^{-15}$ included, \textit{i.e.}
up to remainders $\calO(c^{-19})$, $\calO(c^{-18})$ and
$\calO(c^{-17})$ respectively. We first write the metric components in
the new gauge obtained in Eqs.~\eqref{hI2p}--\eqref{hJ3p} up to the
required order, with the help of Eq.~\eqref{defZP}. For convenience we
simply denote \textit{e.g.}
$\delta{h}_{(1)}^{\mu\nu}=({h'}^{\mu\nu})_{M \times M \times I_{ij}}$,
forgetting about the prime indicating the new gauge and also about the
type of multipole interaction. However we call this piece
$\delta{h}_{(1)}^{\mu\nu}$ because we shall eventually obtain iterated
and twice-iterated contributions $\delta{h}_{(2)}^{\mu\nu}$ and
$\delta{h}_{(3)}^{\mu\nu}$.
\begin{itemize}
\item Mass quadupole:
\begin{subequations}\label{deltahI2}
\begin{align}
\delta{h}_{(1)}^{00} &=
-\frac{1712}{525}\frac{G^3M^2}{c^{13}}\,\hat{x}_{ab}
\int_0^{+\infty}\ud\,\tau\ln\tau
\,\Bigl[I_{ab}^{(8)}(t-\tau)\nonumber\\&\qquad\qquad\qquad
  +\frac{r^2}{14c^2}
  I_{ab}^{(10)}(t-\tau)+\frac{r^4}{504c^4}I_{ab}^{(12)}(t-\tau)\Bigr]
+ \calO\left(\frac{1}{c^{19}}\right)\,,\label{deltahI200}\\
\delta{h}_{(1)}^{0i} &=
\frac{1712}{2205}\frac{G^3M^2}{c^{14}}\,\hat{x}_{iab}\int_0^{+\infty}\ud\,
\tau\ln\tau \,\Bigl[I_{ab}^{(9)}(t-\tau)
  +\frac{r^2}{18c^2}I_{ab}^{(11)}(t-\tau)\Bigr] +
\calO\left(\frac{1}{c^{18}}\right)\,,\\
\delta{h}_{(1)}^{ij} &=
-\frac{428}{2835}\frac{G^3M^2}{c^{15}}\,\hat{x}_{ijab}\int_0^{+\infty}\ud
\,\tau\ln\tau \,I_{ab}^{(10)}(t-\tau) +
\calO\left(\frac{1}{c^{17}}\right)\,.
\end{align}
\end{subequations}
\item Mass octupole:
\begin{subequations}\label{deltahI3}
\begin{align}
\delta{h}_{(1)}^{00} &=
\frac{208}{3969}\frac{G^3M^2}{c^{15}}\,\hat{x}_{abc}
\int_0^{+\infty}\ud\,\tau\ln\tau \,\Bigl[I_{abc}^{(10)}(t-\tau)
  +\frac{r^2}{18c^2} I_{abc}^{(12)}(t-\tau)\Bigr] +
\calO\left(\frac{1}{c^{19}}\right)\,,\\
\delta{h}_{(1)}^{0i} &=
-\frac{52}{5103}\frac{G^3M^2}{c^{16}}\,\hat{x}_{iabc}\int_0^{+\infty}\ud\,
\tau\ln\tau \,I_{abc}^{(11)}(t-\tau) +
\calO\left(\frac{1}{c^{18}}\right)\,,\\
\delta{h}_{(1)}^{ij} &= \calO\left(\frac{1}{c^{17}}\right)\,.
\end{align}
\end{subequations}
\item Mass hexadecapole:
\begin{subequations}\label{deltahI4}
\begin{align}
\delta{h}_{(1)}^{00} &=
-\frac{1571}{1964655}\frac{G^3M^2}{c^{17}}\,\hat{x}_{abcd}
\int_0^{+\infty}\ud\,\tau\ln\tau \,I_{abcd}^{(12)}(t-\tau) +
\calO\left(\frac{1}{c^{19}}\right)\,,\\
\delta{h}_{(1)}^{0i} &= \calO\left(\frac{1}{c^{18}}\right)\,,\\
\delta{h}_{(1)}^{ij} &= \calO\left(\frac{1}{c^{19}}\right)\,.
\end{align}
\end{subequations}
\item Current quadrupole:
\begin{subequations}\label{deltahJ2}
\begin{align}
\delta{h}_{(1)}^{00} &= 0\,,\\
\delta{h}_{(1)}^{0i} &=
\frac{6848}{4725}\frac{G^3M^2}{c^{14}}\,\varepsilon_{iab}\hat{x}_{bc}
\int_0^{+\infty}\ud\,\tau\ln\tau \,\Bigl[J_{ac}^{(8)}(t-\tau) +
  \frac{r^2}{14c^2} J_{ac}^{(10)}(t-\tau)\Bigr]
+\calO\left(\frac{1}{c^{18}}\right)\,,\\
\delta{h}_{(1)}^{ij} &= -
\frac{3424}{6615}\frac{G^3M^2}{c^{15}}\,\varepsilon_{ab(i}\hat{x}_{j)bc}
\int_0^{+\infty}\ud\,\tau\ln\tau \,J_{ac}^{(9)}(t-\tau) +
\calO\left(\frac{1}{c^{17}}\right)\,.
\end{align}
\end{subequations}
\item Current octupole:
\begin{subequations}\label{deltahJ3}
\begin{align}
\delta{h}_{(1)}^{00} &= 0\,,\\
\delta{h}_{(1)}^{0i} &=
-\frac{13}{441}\frac{G^3M^2}{c^{16}}\,\varepsilon_{iab}\hat{x}_{bcd}
\int_0^{+\infty}\ud\,\tau\ln\tau \,J_{acd}^{(10)}(t-\tau)
+\calO\left(\frac{1}{c^{18}}\right)\,,\\
\delta{h}_{(1)}^{ij} &= \calO\left(\frac{1}{c^{17}}\right)\,.
\end{align}
\end{subequations}
\end{itemize}

\subsection{Quadratic iteration}
\label{sec:quad}

At quadratic non-linear order we have to solve the equation
\begin{equation}\label{quadratic}
\Box {h}^{\mu\nu}_{(2)} = {N}^{\mu\nu}_{(2)}\,,
\end{equation}
where the source term ${N}^{\mu\nu}_{(2)}$ is made of quadratic
products of derivatives of ${h}^{\mu\nu}_{(1)}$, symbolically written
as $\sim\partial{h}_{(1)}\partial{h}_{(1)}$ and
$\sim{h}_{(1)}\partial^2{h}_{(1)}$. Here ${h}^{\mu\nu}_{(1)}$ is
composed by the 1PN metric~\eqref{h1PN} augmented at high orders by
all the previous tail-of-tail pieces. Note that Eq.~\eqref{quadratic}
is valid in the new gauge but with the assumption that at the next
non-linear order the harmonic gauge condition is satisfied,
\textit{i.e.} $\partial_\nu{h}^{\mu\nu}_{(2)}=0$, and later, at still
higher order, we shall assume the same,
$\partial_\nu{h}^{\mu\nu}_{(3)}=0$ (such choices are simply a matter of
convenience). The quadratic terms we need for the present iteration,
consistent with the order 7.5PN, are
\begin{subequations}\label{N2munu}
\begin{align}
{N}_{(2)}^{00} + {N}_{(2)}^{ii} &=
-{h}_{(1)}^{00}\partial_{00}{h}_{(1)}^{00} -
2{h}_{(1)}^{0i}\partial_{0i}{h}_{(1)}^{00} -
{h}_{(1)}^{ij}\partial_{ij}{h}_{(1)}^{00} \nonumber\\ & -\partial_i
{h}_{(1)}^{00}\partial_i {h}_{(1)}^{00}-\frac{1}{2}(\partial_0
{h}_{(1)}^{00})^2 +2 \partial_0 {h}_{(1)}^{0i}\partial_0
{h}_{(1)}^{0i}\nonumber\\ & +4\partial_0 {h}_{(1)}^{ij}\partial_i
{h}_{(1)}^{0j}+2\partial_i {h}_{(1)}^{0j}\partial_j
{h}_{(1)}^{0i}+\partial_i {h}_{(1)}^{jk}\partial_i
{h}_{(1)}^{jk}\,,\label{N200}\\
{N}_{(2)}^{0i} &= \frac{3}{4}\partial_0 {h}_{(1)}^{00}\partial_i
{h}_{(1)}^{00} +\partial_j {h}_{(1)}^{00}\partial_i
{h}_{(1)}^{0j}-\partial_j {h}_{(1)}^{00}\partial_j
{h}_{(1)}^{0i}\,,\\
{N}_{(2)}^{ij} &= \frac{1}{4}\partial_i {h}_{(1)}^{00}\partial_j
{h}_{(1)}^{00}-\frac{1}{8}\delta_{ij}\partial_k
{h}_{(1)}^{00}\partial_k {h}_{(1)}^{00}\,.
\end{align}
\end{subequations}
Since we are ultimately interested in the covariant metric components
$g_{\mu\nu}$ we are considering the combination $00 + ii$ of gothic
metric components which appears dominantly into $g_{00}$.

We now replace in~\eqref{N2munu} the gothic metric by its explicit form which
reduces up to 1PN order to Eqs.~\eqref{h1PN} and involves all the tail-of-tail
pieces $\delta{h}_{(1)}^{\mu\nu}$. Obviously the iterated quadratic
tail-of-tail pieces will come from the cross products between the 1PN
metric~\eqref{h1PN} and the linear
tails of tails~\eqref{deltahI2}--\eqref{deltahJ3}. When considering such cross
products, we shall have to integrate typical terms whose general structure is
$\hat{x}_L\phi$, where $\phi$ is any of the potentials appearing in
Eqs.~\eqref{h1PN}, and where $\hat{x}_L$ denotes a STF product of spatial
vectors coming from Eqs.~\eqref{deltahI2}--\eqref{deltahJ3}. Notice that the
hereditary integrals therein are simply functions of time [see
\textit{e.g.}~\eqref{Fij}], and essentially play a spectator role in the
process, with the notable exception of that in the dominant term for which we
have to consider a retardation at the relative 1PN order. In addition, because
of the 1PN retardation in $\delta{h}^{\mu\nu}_{(1)}$, we shall have to
integrate the slightly more complicated source term $r^2\hat{x}_L\phi$ [see
\textit{e.g.}~Eq. \eqref{deltahI200}]. Thus, the equations we have to solve
are
\begin{subequations}\label{twoeqs}
\begin{align}
\Delta \Psi_L &= \hat{x}_L\,\phi\,,\label{eq1}\\ \Delta \Phi_L &=
r^2\hat{x}_L\,\phi\,.\label{eq2}
\end{align}
\end{subequations}
To solve them we adopt the method of ``super-potentials''. Namely we
introduce, given the potential $\phi$, the hierarchy of its
super-potentials denoted $\phi_{2k+2}$, for any positive integer $k$,
where $\phi_{0}=\phi$ and
\begin{equation}
\Delta \phi_{2k+2} = \phi_{2k}\,.
\end{equation}
We thus have $\Delta^k\phi_{2k}=\phi$. The explicit formulas for the
solutions of Eqs.~\eqref{twoeqs} (the first one being needed for
$\ell=4$, and the second one only for $\ell=0,1$) are
\begin{subequations}\label{twosol}
\begin{align}
\Psi_L = \Delta^{-1} \bigl(\hat{x}_{L}\,\phi\bigr) &=
\sum_{k=0}^{\ell}\frac{(-2)^k\ell!}{(\ell-k)!}\,x^{\langle
  L-K}\partial_{K\rangle}\phi_{2k+2}\,,\label{sol1}\\ \Phi_L =
\Delta^{-1} \bigl(r^2\hat{x}_{L}\,\phi\bigr) &=
\sum_{k=0}^{\ell}\frac{(-2)^k\ell!}{(\ell-k)!}\,x^{\langle
  L-K}\partial_{K\rangle} \Bigl[r^2\phi_{2k+2}\nonumber\\&
\qquad \qquad +2(k+1)(2k+1)
  \phi_{2k+4}-4(k+1)x^i\partial_i\phi_{2k+4}\Bigr]\,.\label{sol2}
\end{align}
\end{subequations}
These solutions are unique in the following sense. Suppose that $\phi$
admits an asymptotic expansion when $r\to\infty$ (with $t$ fixed) on
the set of basis functions $r^{\lambda-n}$, labeled by
$n\in\mathbb{N}$ and where the maximal power is
$\lambda\in\mathbb{R}\!\setminus\!\mathbb{N}$ (\textit{i.e.}, is not
an integer). Then, for instance, the solution $\Psi_L$ given by
Eq.~\eqref{sol1} is the unique solution of Eq.~\eqref{eq1}, valid
\textit{in the sense of distribution theory}~\cite{Schwartz}, that
admits an asymptotic expansion when $r\to\infty$ on the basis
functions $r^{\lambda+\ell+2-n}$. Similarly $\Phi_L$ is the unique
solution in the sense of distributions which admits an asymptotic
expansion on the basis $r^{\lambda+\ell+4-n}$. The
formulas~\eqref{twosol} can be easily proved by induction. They can
also be iterated if necessary; for instance we find by iterating $i$
times the first one that
\begin{equation}\label{sol3}
\Delta^{-i} \Psi_L = \Delta^{-i-1} \bigl(\hat{x}_{L}\,\phi\bigr) =
\sum_{k=0}^{\ell}\frac{(-2)^k(k+i)!\ell!}{k!i!(\ell-k)!}\,x^{\langle
  L-K}\partial_{K\rangle}\phi_{2k+2i+2}\,.
\end{equation}

Let us give an example of the applicability of those formulas. The 1PN
compact-support potential $U$ was defined by Eq.~\eqref{U}, where we
recall that the effective masses $\tilde{\mu}_A$ are mere functions of
time. Now the hierarchy of super-potentials of $U$ is given by
\begin{equation}\label{U2k}
U_{2k}=\frac{1}{(2k)!}\Bigl[G \tilde{\mu}_1\,r_1^{2k-1} + G
  \tilde{\mu}_2\,r_2^{2k-1}\Bigr]\,.
\end{equation}
For $k=1$ we recover the potential $U_2$ already met in
Eq.~\eqref{U2}. Of course similar expressions apply for the other
potentials $U_i$ and $U_{ij}$ in Eqs.~\eqref{potcompact}. Taking only
the leading-order cross term in the expression of the non-linear
source~\eqref{N200} we find that we have to solve 
\begin{equation}\label{BoxPsi}
\Box \Psi = x_i \,\partial_j V \,F_{ij}(t)\,.
\end{equation}
Here $V$ is the retarded potential~\eqref{V} and $F_{ij}$ is a certain
function of time, which we shall define below to be the hereditary
integral~\eqref{Fij}. Note that Eq.~\eqref{BoxPsi} is to be solved
including the first-order retardation at 1PN order, which is simply
done by using the symmetric propagator
$\Box^{-1}=\Delta^{-1}+\frac{1}{c^2}\partial_t^2\Delta^{-2}+\calO(c^{-2})$.
Using then our elementary solution~\eqref{sol1} we get, up to 1PN
relative order,
\begin{align}\label{Psi1PN}
\Psi = \Box^{-1}\Bigl[x_i \,\partial_j V \,F_{ij}(t)\Bigr] &=
\left(x_i\partial_j U_{2}-2\partial_{ij}U_{4}\right)F_{ij}\\ &+
\frac{1}{c^{2}}\biggl[ x_i\left(2\partial_j\partial_t^2
  U_{4}F_{ij}+2\partial_{j}\partial_tU_{4}F^{(1)}_{ij} +
  \partial_{j}U_{4}F^{(2)}_{ij}\right) \nonumber\\ &\quad\quad - 6
  \partial_{ij}\partial_t^2U_{6}F_{ij} - 8
  \partial_{ij}\partial_tU_{6}F^{(1)}_{ij} - 4
  \partial_{ij}U_{6}F^{(2)}_{ij} \biggr] +
\calO\left(\frac{1}{c^{4}}\right)\,.\nonumber
\end{align} 

Below we shall need not only the super-potentials of a compact-support
potential like $U$, but also those of more complicated potentials such
as $\hat{W}_{ij}$ defined by Eq.~\eqref{Wij}. Its first order
super-potential reads (to Newtonian order)
\begin{subequations}\label{Wij2}
\begin{align}
\hat{W}_2^{ij} &= U_2^{ij}-\delta^{ij}U_2^{kk}\nonumber\\ &
-\frac{G^2m_1^2}{8}\left[\partial_{ij}\left(\frac{r_1^2}{6}\Bigl(\ln
  r_1-\frac{5}{6}\Bigr)\right)+\delta^{ij}\ln
  r_1\right]-\frac{G^2m_2^2}{8}\left[\partial_{ij}\left(\frac{r_2^2}{6}\Bigl(\ln
  r_2-\frac{5}{6}\Bigr)\right)+\delta^{ij}\ln r_2\right]\nonumber\\ &
-G^2m_1m_2\frac{\partial^2f}{\partial y_1^{(i}\partial y_2^{j)}} \,,
\end{align}
\end{subequations}
where we have used the super-potential of $U_{ij}$ as well as the one
of the function $g$ of Eq.~\eqref{g}, namely $g_2=f/2$ defined by
\begin{equation}\label{f}
f = \frac{1}{3} \,\mathbf{r}_1\cdot\mathbf{r}_2\Bigl[g-\frac{1}{3}\Bigr]
+ \frac{1}{6}\left(r_1r_{12}+r_2r_{12}-r_1r_2\right)\,,
\end{equation}
where $\mathbf{r}_A=\mathbf{x}-\mathbf{y}_A$, which satisfies $\Delta
f = 2g$ in the sense of distributions (see \textit{e.g.}
Ref.~\cite{BFeom}). A full hierarchy of higher super-potentials for
the function $g$ could be defined similarly. Note that the
super-potentials of the non-compact potential $U^2$ are obtained
thanks to the super-potentials of $g$ (at Newtonian order say,
\textit{i.e.} assimilating $\tilde{\mu}_A$ to $m_A$):
\begin{subequations}\label{superg}
\begin{align}
U^2 &= \frac{G^2m_1^2}{r_1^2} + \frac{G^2m_2^2}{r_2^2} + 2
\frac{G^2m_1m_2}{r_1 r_2}\,,\\ \bigl(U^2)_2 &= G^2m_1^2 \ln r_1 +
G^2m_2^2 \ln r_2 + 2 G^2m_1m_2 g\,,\\ \bigl(U^2)_4 &=
\frac{G^2m_1^2}{6} r_1^2 \Bigl(\ln r_1-\frac{5}{6}\Bigr) +
\frac{G^2m_2^2}{6} r_2^2 \Bigl(\ln r_2-\frac{5}{6}\Bigr) + G^2m_1m_2
f\,.
\end{align} 
\end{subequations}

We now define as a convenient short-hand the following hereditary
function of time appropriate for the mass quadrupole moment,
\begin{equation}\label{Fij}
F_{ij}(t) = - \frac{1712}{525} G^3 M^2
\int_0^{+\infty}\ud\,\tau\ln\tau \,I_{ij}^{(8)}(t-\tau)\,,
\end{equation}
and obtain the full expressions of $\delta{h}_{(2)}^{\mu\nu}$ up to
the requested PN order as
\begin{subequations}\label{h2I2}
\begin{align}
\delta{h}_{(2)}^{00} + \delta{h}_{(2)}^{ii} &=
\frac{16}{c^{15}}\left(x_i\partial_j
U_{2}-2\partial_{ij}U_{4}\right)F_{ij} \nonumber\\ &+
\frac{1}{c^{17}}\biggl[ 16 x_i\left(2\partial_i\partial_t^2
  U_{4}F_{ij}+2\partial_{i}\partial_tU_{4}F^{(1)}_{ij}+\partial_{i}U_{4}F^{(2)}_{ij}\right)
  \nonumber\\ &\qquad\quad +16\left(- 6 \partial_{ij}\partial_t^2
  U_{6}F_{ij} - 8 \partial_{ij}\partial_tU_{6}F^{(1)}_{ij} - 4
  \partial_{ij}U_{6}F^{(2)}_{ij}\right)
  \nonumber\\ &\qquad\quad+\frac{8}{5}\left(r^2x_i\partial_j
  U_{2}-2x_i\partial_{j}U_{4}-4x_{ik}\partial_{kj}U_4-2r^2\partial_{ij}U_4\right.\nonumber\\ &\qquad\qquad\quad\left.-8\partial_{ij}U_6+16x_k\partial_{ijk}U_6\right)F^{(2)}_{ij}
  \nonumber\\ &\qquad\quad+\frac{8}{7}\left(\hat{x}_{ijk}\partial_k
  U_{2}-6x_{\langle ij}\partial_{k\rangle}\partial_kU_{4}+24x_{\langle
    i}\partial_{jk\rangle}\partial_kU_6
  -48\hat{\partial}_{ijk}\partial_k U_8\right)F^{(2)}_{ij}
  \nonumber\\ &\qquad\quad+4\left(\hat{x}_{ij} U_{2}-4x_{\langle
    i}\partial_{j\rangle}U_{4}+8\partial_{ij}U_6\right)F^{(2)}_{ij}
  \nonumber\\ &\qquad\quad+4\left(\hat{x}_{ij}
  \partial^2_tU_{2}-4x_{\langle i}\partial_{j\rangle}\partial^2_tU_{4}
  +8\partial_{ij}\partial^2_tU_6\right)F_{ij}
  \nonumber\\ &\qquad\quad+16\left(\hat{x}_{i}
  U^j_{2}-2\partial_{i}U^j_{4}\right)F^{(1)}_{ij}
  \nonumber\\ &\qquad\quad-\frac{40}{21}\left(\hat{x}_{ijk}\partial_k
  \partial_t U_{2}-6x_{\langle
    ij}\partial_{k\rangle}\partial_k\partial_tU_{4}+24x_{\langle
    i}\partial_{jk\rangle}\partial_k\partial_tU_6\right.\nonumber\\ &\qquad\qquad\quad\left.-
  48\hat{\partial}_{ijk}\partial_k\partial_t U_8\right)F^{(1)}_{ij}
  \nonumber\\ &\qquad\quad +
  \frac{5}{27}\left(\hat{x}_{ijkl}\partial_{kl} U_{2}-8x_{\langle
    ijk}\partial_{l\rangle}\partial_{kl}U_{4}+48x_{\langle
    ij}\partial_{kl\rangle}\partial_{kl}U_6\right.\nonumber\\ &\qquad\qquad\quad\left.-192x_{\langle
    i}\partial_{jkl\rangle}\partial_{kl}U_8
  +384\hat{\partial}_{ijkl}\partial_{kl}U_{10}\right)F^{(2)}_{ij}
  \nonumber\\ &\qquad\quad+4\left(\hat{x}_{ij}
  \partial_tU_{2}-4x_{\langle i}\partial_{j\rangle}\partial_tU_{4}
  +8\partial_{ij}\partial_tU_6\right)F^{(1)}_{ij}
  \nonumber\\ &\qquad\quad+\frac{80}{21}\left(\hat{x}_{ij}
  \partial_kU^k_{2}-4x_{\langle i}\partial_{j\rangle}\partial_kU^k_{4}
  +8\hat{\partial}_{ij}\partial_kU^k_6\right)F^{(1)}_{ij}
  \nonumber\\ &\qquad\quad+\frac{160}{21}\left(\hat{x}_{ik}
  \partial_kU^j_{2}-4x_{\langle i}\partial_{k\rangle}\partial_kU^j_{4}
  +8\hat{\partial}_{ik}\partial_kU^j_6\right)F^{(1)}_{ij}
  \nonumber\\ &\qquad\quad-\frac{64}{21}\left(\hat{x}_{ik}
  \partial_jU^k_{2}-4x_{\langle i}\partial_{k\rangle}\partial_jU^k_{4}
  + 8\hat{\partial}_{ik}\partial_jU^k_6\right)F^{(1)}_{ij}
  \nonumber\\ &\qquad\quad+ 8 \left(\hat{W}_2^{ij} +
  x_i\partial_j\bigl(\hat{W}+4U^2\bigr)_2 -
  2\partial_{ij}\bigl(\hat{W}+4U^2\bigr)_4\right)F_{ij}\biggr] +
\calO\left(\frac{1}{c^{19}}\right)\,,\\
\delta{h}_{(2)}^{0i} &= \frac{1}{c^{16}}\biggl[-6\left(x_j\partial_t
  U_2-2\partial_{j}\partial_t U_4\right)F_{ij} \nonumber\\&\qquad\quad
  - 3\left(x_{jk}\partial_i U_2-4x_{j}\partial_{ik}
  U_4+8\partial_{ijk}U_6\right)F^{(1)}_{jk} \nonumber\\&\qquad\quad +
  \frac{8}{3}\left(\hat{x}_{jk}\partial_k U_2-4x_{\langle
    j}\partial_{k\rangle}\partial_{k}
  U_4+8\hat{\partial}_{jk}\partial_kU_6\right)F^{(1)}_{ij}\nonumber\\&\qquad\quad
  - \frac{8}{3}\left(\hat{x}_{ij}\partial_k U_2-4x_{\langle
    i}\partial_{j\rangle}\partial_{k}
  U_4+8\hat{\partial}_{ij}\partial_kU_6\right)F^{(1)}_{jk}\nonumber\\&\qquad\quad
  - 8 \left(x_{j}\left(\partial_i U^k_2-\partial_k
  U^i_2\right)-2\left(\partial_{ij}U^k_4-\partial_{jk}U^i_4\right)\right)F_{jk}\biggr]+
\calO\left(\frac{1}{c^{18}}\right)\,,\\
\delta{h}_{(2)}^{ij} &= \frac{1}{c^{15}}\biggl[
  -4\left(x_k\partial_{(i}U_2-2\partial_{k(i}U_4\right)F_{j)k}\nonumber\\&
  \qquad\quad
  +2\delta_{ij}\left(x_k\partial_{l}U_2-2\partial_{kl}U_4\right)F_{kl}
  \biggr] + \calO\left(\frac{1}{c^{17}}\right)\,.
\end{align}
\end{subequations}
We must also do the same for the other multipole interactions, but
these arise at higher PN order and the iteration is much simpler. We
need only to consider the mass octupole and current quadrupole
moments, for which we define
\begin{subequations}\label{GH}
\begin{align}
G_{ijk}(t) &= \frac{208}{3969} G^3 M^2
\int_0^{+\infty}\ud\,\tau\ln\tau
\,I_{ijk}^{(10)}(t-\tau)\,,\\ H_{ij}(t) &= \frac{6848}{4725} G^3 M^2
\int_0^{+\infty}\ud\,\tau\ln\tau \,J_{ij}^{(8)}(t-\tau)\,.
\end{align}
\end{subequations}
For the mass octupole moment we obtain
\begin{equation}\label{h2I3}
\delta{h}_{(2)}^{00} + \delta{h}_{(2)}^{ii} = \frac{24}{c^{17}}\biggl[
  x_{ij}\partial_k U_{2} - 4 x_{i}\partial_{jk} U_{4} +
  8\partial_{ijk}U_{6}\biggr] G_{ijk} +
\calO\left(\frac{1}{c^{19}}\right)\,,
\end{equation}
while the other components are negligible. For the current quadrupole
we have
\begin{subequations}\label{h2J2}
\begin{align}
\delta{h}_{(2)}^{00} + \delta{h}_{(2)}^{ii} &= \frac{8}{c^{17}}\biggl[
  \left( \hat{x}_{bc}\partial_i \partial_t U_{2} - 4 x_{\langle
    b}\partial_{c\rangle}\partial_i \partial_t U_{4} +
  8\hat{\partial}_{bc} \partial_i \partial_t U_{6}\right)
  \varepsilon_{iab}H_{ac} \nonumber\\ &\qquad\quad -
  \frac{5}{28}\left(\hat{x}_{jbc}\partial_{ij} U_{2} - 6
  \hat{x}_{\langle jb}\partial_{c\rangle}\partial_{ij} U_{4} +
  24x_{\langle j}\partial_{bc\rangle}\partial_{ij}U_6 - 48
  \hat{\partial}_{jbc} \partial_{ij}
  U_8\right)\varepsilon_{iab}H^{(1)}_{ac}\nonumber\\ &\qquad\quad +
  2\left(x_c\partial_iU^j_2
  -2\partial_{ci}U^j_4\right)\varepsilon_{ija}H_{ac} -
  2\left(x_b\partial_iU^j_2
  -2\partial_{bi}U^j_4\right)\varepsilon_{iab}H_{aj}\biggr]
\nonumber\\ &+ \calO\left(\frac{1}{c^{19}}\right)\,,\\
\delta{h}_{(2)}^{0i} &= \frac{4}{c^{16}}\biggl[ -2
  \left(x_{c}\partial_j U_{2} - 2\partial_{cj} U_{4} \right)
  \varepsilon_{ija}H_{ac} \nonumber\\ &\qquad\quad +
  \left(x_{b}\partial_j U_{2} - 2\partial_{bj}
  U_{4}\right)\bigl(\varepsilon_{iab}H_{aj}-\varepsilon_{jab}H_{ai}\bigr)
  \biggr]+ \calO\left(\frac{1}{c^{18}}\right)\,,
\end{align}
\end{subequations}
while the $ij$ components are negligible.

\subsection{Cubic iteration}
\label{sec:cub}

At the next-to-next-to-leading 7.5PN order (\textit{i.e.} 2PN
relative) it is evident that there is one further iteration to be
performed. However that iteration will concern only the mass
quadrupole moment that appears at the leading 5.5PN order. For that
moment we have to integrate the cubic equation
\begin{equation}\label{cubic}
\Box {h}^{\mu\nu}_{(3)} = {M}^{\mu\nu}_{(3)} + {N}^{\mu\nu}_{(3)}\,.
\end{equation}
The cubic source term is the sum of two contributions:
${M}^{\mu\nu}_{(3)}$ which is a direct product of three linear terms
${h}^{\mu\nu}_{(1)}$ and can be symbolically written as
$\sim{h}_{(1)}\partial{h}_{(1)}\partial{h}_{(1)}$, and
${N}^{\mu\nu}_{(3)}$ which is a product between a linear term
${h}^{\mu\nu}_{(1)}$ and a quadratic one ${h}^{\mu\nu}_{(2)}$,
symbolically written as $\sim\partial{h}_{(1)}\partial{h}_{(2)}$. At
cubic order only the dominant contribution in the combination $00+ii$
of the components of the source terms will be needed.

Considering first the $M^{\mu\nu}_{(3)}$ piece we find that the
dominant contribution therein is
\begin{equation}\label{M3dom}
{M}_{(3)}^{00} + {M}_{(3)}^{ii} = -
\frac{9}{8}\,{h}_{(1)}^{00}\partial_i{h}_{(1)}^{00}\partial_i{h}_{(1)}^{00}
\,.
\end{equation}
We replace the linear metric ${h}_{(1)}^{00}$ by its explicit
expression made of the sum of Eqs.~\eqref{h1PN00}
and~\eqref{deltahI200} (in which only the leading term of order
$c^{-13}$ is to be included), and find again that the integration can
be explicitly performed thanks to the method of
super-potentials. However we need to compute the super-potentials of
slightly more complicated potentials with non-compact support. A first
series is (with $U$ Newtonian)
\begin{subequations}\label{super1}
\begin{align}
\bigl(U\partial_iU\bigr)_2 &= \frac{G^2m_1^2}{2}\partial_{i}\ln r_1 +
\frac{G^2m_2^2}{2}\partial_{i}\ln r_2 - G^2m_1m_2\left(\frac{\partial
  g}{\partial y_1^{i}} +\frac{\partial g}{\partial
  y_2^{i}}\right)\,,\\ \bigl(U\partial_iU\bigr)_4 &=
\frac{G^2m_1^2}{12}\partial_{i}\left[r_1^2\left(\ln
  r_1-\frac{5}{6}\right)\right] +
\frac{G^2m_2^2}{12}\partial_{i}\left[r_2^2\left(\ln
  r_2-\frac{5}{6}\right)\right] \nonumber\\&-
\frac{G^2m_1m_2}{2}\left(\frac{\partial f}{\partial y_1^{i}}
+\frac{\partial f}{\partial y_2^{i}}\right)\,,
\end{align} 
\end{subequations}
where $f$ has been defined by Eq.~\eqref{f}. To define another series
we introduce the super-potential of $U\Delta U$ namely (at Newtonian
order)
\begin{equation}\label{K}
K = \left(U\Delta U\right)_2 = \frac{G m_1}{r_1}(U)_1+\frac{G
  m_2}{r_2}(U)_2\,,
\end{equation}
with $(U)_1=G m_2/r_{12}$ and $(U)_2=G m_1/r_{12}$ being the values of $U$ at the
locations of the particles. Notice that in fact (at Newtonian order) $K$ is
related to the trace $\hat{W}=\hat{W}_{kk}$ of Eq.~\eqref{Wij} by
\begin{equation}\label{KW}
K = \hat{W} + \frac{U^2}{2} + 2 U_{kk}\,.
\end{equation}
Then we can write, with the super-potentials of $K$ computed
similarly to Eq.~\eqref{U2k},
\begin{subequations}\label{super2}
\begin{align}
\bigl(\partial_iU\partial_iU\bigr)_2 &= - K +
\frac{U^2}{2}\,,\\ \bigl(\partial_iU\partial_iU\bigr)_4 &= - K_2 +
\frac{G^2m_1^2}{2} \ln r_1 + \frac{G^2m_2^2}{2} \ln r_2 + G^2m_1m_2\,g
\,,\\ \bigl(\partial_iU\partial_iU\bigr)_6 &= - K_4 +
\frac{G^2m_1^2}{12} r_1^2\left(\ln r_1-\frac{5}{6}\right) +
\frac{G^2m_2^2}{12} r_2^2\left(\ln r_2-\frac{5}{6}\right) +
\frac{G^2m_1m_2}{2}\,f\,.
\end{align} 
\end{subequations}
With these results we obtain in fully closed form the solution
corresponding to the direct cubic source term~\eqref{M3dom} as
\begin{align}\label{h3part1}
\delta{h}_{(3)}^{00} + \delta{h}_{(3)}^{ii} &=
\frac{18}{c^{17}}\biggl[-4x_{i}\bigl(U\partial_jU\bigr)_2 +
  8\partial_i\bigl(U\partial_jU\bigr)_4 -
  x_{ij}\bigl(\partial_kU\partial_kU\bigr)_2 \nonumber\\&\quad\quad +
  4x_{i}\partial_j\bigl(\partial_kU\partial_kU\bigr)_4 -
  8\partial_{ij}\bigl(\partial_kU\partial_kU\bigr)_6\biggr] F_{ij}+
\calO\left(\frac{1}{c^{19}}\right)\,,
\end{align}
with, as we said, the other components $0i$ and $ij$ being negligible
at this stage.

Considering next the ${N}^{\mu\nu}_{(3)}$ piece of the cubic source
term~\eqref{cubic} we find that only the following contributions are
needed:
\begin{equation}\label{cubicN}
{N}_{(3)}^{00} + {N}_{(3)}^{ii} = - {h}_{(1)}^{ij}\partial_{ij}
{h}_{(2)}^{00} - {h}_{(2)}^{ij}\partial_{ij} {h}_{(1)}^{00} -2
\partial_i {h}_{(1)}^{00}\partial_i {h}_{(2)}^{00}\,.
\end{equation}
Again the method of super-potentials works for all the terms
encountered. The needed super-potentials are some straightforward
extensions or variants of the ones in Eqs.~\eqref{super1}
and~\eqref{super2}. Let us add that the super-potentials of $r_1/r_2$
and $r_2/r_1$ are also needed. These are obtained by appropriate
exchanges between the field point $\mathbf{x}$ and the source points
$\mathbf{y}_A$ in Eq.~\eqref{f}. Posing
\begin{subequations}\label{f12}
\begin{align}
f_{12} &=
-\frac{1}{3}\mathbf{r}_1\cdot\mathbf{r}_{12}\Bigl[g-\frac{1}{3}\Bigr]
+ \frac{1}{6}\left(r_1r_2+r_2r_{12}-r_1r_{12}\right)\,,\\f_{21} &=
\frac{1}{3}\mathbf{r}_2\cdot\mathbf{r}_{12}\Bigl[g-\frac{1}{3}\Bigr] +
\frac{1}{6}\left(r_1r_2+r_1r_{12}-r_2r_{12}\right)\,,
\end{align}
\end{subequations}
we have $\Delta f_{12}=r_1/r_2$ and $\Delta f_{21}=r_2/r_1$. Those
solutions appear in the more complicated super-potential
\begin{subequations}\label{super3}
\begin{align}
\bigl(U_2\,\partial_{ij}U\bigr)_2 =& - \frac{G^2m_1^2}{8}
\left[\partial_{ij}\left(r_1^2\left(\ln r_1-\frac{5}{6}\right)\right)
  -2\delta_{ij}\ln r_1\right] \nonumber\\ & - \frac{G^2m_2^2}{8}
\left[\partial_{ij}\left(r_2^2\left(\ln r_2-\frac{5}{6}\right)\right)
  -2\delta_{ij}\ln r_2\right]\nonumber\\ & +
\frac{G^2m_1m_2}{2}\left(\frac{\partial^2 f_{21}}{\partial
  y_1^{ij}}+\frac{\partial^2 f_{12}}{\partial y_2^{ij}}\right)\,.
\end{align} 
\end{subequations}
Finally we encounter a series of super-potentials with compact support
generalizing Eq.~\eqref{K}, of the type
\begin{equation}\label{potphi2k}
\left(\phi\Delta U\right)_{2k+2}=\frac{1}{(2k)!}\Bigl[G m_1\,(\phi)_1
  r_1^{2k-1} + G m_2\,(\phi)_2 r_2^{2k-1}\Bigr]\,.
\end{equation}
where $(\phi)_A$ denotes the value of $\phi$ at the particle $A$. We
are finally in a position to write down the complete explicit form of
the cubic solution of Eq.~\eqref{cubicN} as
\begin{align}\label{h3part2}
\delta{h}_{(3)}^{00} + \delta{h}_{(3)}^{ii} &=
\frac{8}{c^{17}}\biggl[x_{i}\left(7U\partial_jU_2-U_2\partial_jU\right)
  - 14U\partial_{ij}U_4+2\partial_iU\partial_jU_4
  -6x_i\bigl(U\partial_jU\bigr)_2\nonumber\\&\quad\quad
  +12\partial_i\bigl(U\partial_jU\bigr)_4 +
  2\bigl(U_2\partial_{ij}U\bigr)_2 -7x_i\bigl(\partial_jU_2\Delta
  U\bigr)_2+x_i\partial_j\bigl(U_2\Delta U\bigr)_2
  \nonumber\\&\quad\quad+ 14\partial_i\bigl(\partial_jU_2\Delta
  U\bigr)_4 -2\partial_{ij}\bigl(U_2\Delta U\bigr)_4
  +14\bigl(\partial_{ij}U_4\Delta
  U\bigr)_2-2\partial_i\bigl(\partial_jU_4\Delta U\bigr)_2\biggr]
F_{ij} \nonumber\\& + \calO\left(\frac{1}{c^{19}}\right)\,.
\end{align}

\subsection{Miscellaneous}
\label{sec:misc}

A few operations are still in order before obtaining the relevant
metric and the result for the redshift factor~\eqref{uT}. Of course we
have to sum up all the results, thereby obtaining the full (iterated
and twice-iterated) tail-of-tail contributions in the gothic metric
deviation,
\begin{equation}\label{gothictotal}
\delta {h}^{\mu\nu} = \delta {h}^{\mu\nu}_{(1)} + \delta
       {h}^{\mu\nu}_{(2)} + \delta {h}^{\mu\nu}_{(3)}\,,
\end{equation}
where $\delta {h}^{\mu\nu}_{(1)}$ is itself the sum of
Eqs.~\eqref{deltahI2} to~\eqref{deltahJ3}, $\delta {h}^{\mu\nu}_{(2)}$
is the sum of~\eqref{h2I2},~\eqref{h2I3} and~\eqref{h2J2}, and $\delta
{h}^{\mu\nu}_{(3)}$ is the sum of~\eqref{h3part1}
and~\eqref{h3part2}. The corresponding contributions in the usual
covariant metric, say $\delta g_{\mu\nu}$, must then be deduced
from~\eqref{gothictotal}. This is a straightforward step and we get,
up to the requested PN order,
\begin{subequations}\label{gcov}
\begin{align}
\delta g_{00} &=
-\frac{1}{2}\left(1+\frac{h^{00}+h^{ii}}{2}\right)\left(\delta h^{00}+
\delta h^{ii}\right) - \frac{1}{2}h^{00}\delta h^{00}+ h^{0i}\delta
h^{0i} + \frac{1}{2}h^{ij}\delta h^{ij} \nonumber\\ & -
\frac{15}{16}(h^{00})^2\delta h^{00} +
\calO\left(\frac{1}{c^{19}}\right) \,,\\ \delta g_{0i} &=
\left(1+\frac{h^{00}}{2}\right)\delta h^{0i} + \frac{1}{2}
h^{0i}\delta h^{00} + \calO\left(\frac{1}{c^{18}}\right)\,,\\ \delta
g_{ij} &= -\delta h^{ij} + \frac{1}{2}\left(-\delta h^{00}+\delta
h^{kk}\right)\delta_{ij} - \frac{1}{4} h^{00}\delta h^{00} \delta^{ij}
+ \calO\left(\frac{1}{c^{17}}\right)\,,
\end{align}
\end{subequations}
where $h^{\mu\nu}$ is the 1PN gothic metric~\eqref{h1PN}.

Next we have to single out the conservative part of the metric,
\textit{i.e.} neglect the dissipative radiation reaction effects. As
in Paper~I we assume that the split between conservative and
dissipative effects is equivalent to a split between
``time-symmetric'' and ``time-antisymmetric'' contributions in the
following sense. We decompose each of the tail-of-tail integrals like
for instance $F_{ij}$ defined in Eq.~\eqref{Fij}, into
$F_{ij}=F^\text{cons}_{ij}+F^\text{diss}_{ij}$ where
\begin{subequations}\label{Fijcons}
\begin{align}
F^\text{cons}_{ij}(t) = - \frac{1712}{1050} G^3 M^2
\int_0^{+\infty}\ud\,\tau\ln\tau \Bigl[I_{ij}^{(8)}(t-\tau) +
  I_{ij}^{(8)}(t+\tau)\Bigr]\,,\\ F^\text{diss}_{ij}(t) = -
\frac{1712}{1050} G^3 M^2 \int_0^{+\infty}\ud\,\tau\ln\tau
\Bigl[I_{ij}^{(8)}(t-\tau) - I_{ij}^{(8)}(t+\tau)\Bigr]\,,
\end{align}\end{subequations}
and keep only the conservative part that is time-symmetric. This was
justified in Paper~I by the fact that the equations of motion of
compact binaries associated with the conservative part of the metric
defined in that way are indeed conservative, \textit{i.e.} the
acceleration is purely radial for circular orbits.

From the equations of motion reduced to circular orbits we obtain the
relation between the separation $r_{12}$ between the particles and the
orbital frequency $\Omega$. This relation is important when we reduce
the expressions to the frame of the center of mass and then to
circular orbits. We have checked that the results obtained in
Eqs.~(5.16)--(5.17) of Paper~I are sufficient for the present
purpose. However it is important that in all relations (such as the
one between orbital separation and frequency) we take into account the
lowest order 2PN corrections, appropriate when performing a
next-to-next-to-leading computation. For the same reason it is also
important, when we replace the complete covariant metric
$g_{\mu\nu}$ in the redshift factor defined by Eq.~\eqref{uT}, to
include not only all the high order tail-of-tail pieces, but also the
lower order covariant metric up to 2PN order, because of couplings
between the 2PN metric and the various iterated tail-of-tail pieces at
next-to-next-to-leading order. We do not reproduce here the 2PN metric
at the location of each particle since it is given in full form by
Eqs.~(7.6) of Ref.~\cite{BFP98}.
 
\section{Discussion}
\label{sect:disc}

In this paper, using standard post-Newtonian methods (see
\textit{e.g.}~\cite{Bliving14}), we have computed
next-to-next-to-leading contributions to Detweiler's redshift
variable~\cite{Det08} at odd powers in the post-Newtonian expansion,
by examining the conservative post-Newtonian dynamics of compact
binaries moving on exactly circular orbits. Conservative PN effects at
odd powers in the PN expansion necessarily involve non-local in time
or hereditary (tail) integrals extending over the whole past history
of the source~\cite{BFW14a}. They have been shown to appear first at
the 5.5PN order in the redshift factor for circular
orbits~\cite{SFW14}. In the standard PN approximation they have been
proved to originate from the so-called tails of tails associated with
the mass quadrupole moment of the source~\cite{BFW14a}.

Here we have extended our previous effort to 2PN order beyond the
leading 5.5PN contribution, thus obtaining the 6.5PN and 7.5PN
coefficients in the redshift factor (at linear order in the mass
ratio), which are perhaps the highest orders ever reached by
traditional PN methods. This work involved computing high-order
tails of tails associated with higher mass and current multipole
moments. For this purpose, we have systematically worked in a
preferred gauge for which the computation drastically simplifies, with
respect to, say, the harmonic gauge. In addition we have employed a
more efficient method to obtain the precise coefficients of
tail-of-tail integrals in the near zone of general matter
sources. Furthermore, we could perform the non-linear iteration of
tails of tails thanks to an integration method based on the use of
hierarchical ``super-potentials''. Our analytical post-Newtonian
calculation gives results in full agreement with numerical and
analytical self-force calculations~\cite{SFW14, BiniD14b}.

The present work is an addition to the body of works~\cite{Det08,
  BDLW10a, BDLW10b, LBW12, BFW14a} that have demonstrated the
beautiful consistency between analytical post-Newtonian methods, valid
for any matter source but limited to the weak-field slow-motion regime
of the source, and gravitational self-force methods, which give an
accurate description of extreme mass ratio compact binaries even in
the relativistic and strong-field regime. The agreement between PN and
GSF approaches provides an indirect check that the dimensional
regularization procedure invoked in the PN calculation when it is
applied to point particle binary sources, is in fact equivalent to the
very different procedure of subtraction of the singular field which is
employed in the GSF approach. Although the dimensional regularization
has not been explicitly used in the present paper, this check between
very different regularization procedures was a central motivation for
our initial works~\cite{BDLW10a, BDLW10b}. 
Our recent work~\cite{BFW14a} together with the
present paper confirm that the machinery used in the traditional PN
approach to compute non-linear effects and their associated
hereditary-type integrals like tails, tails of tails and so on, is
correct.

In principle, the prospects of extending the present analysis to yet
higher PN orders are good. The main challenge would be to control higher
non-linear multipole interactions. In particular, the computation of the
coefficient at 8.5PN order would be feasible since we know the mass
quadrupole moment to 3PN order. On the other hand, extension to
higher-order in the mass ratio would be possible only by controlling
other multipole couplings such as the double mass-quadrupole interaction
coupled with mass monopoles.

The success of the comparison performed in this paper has obviously important
implications: post-Newtonian calculations of tails of tails at 3PN order
beyond the (2.5PN) quadrupole term already play a role~\cite{Bliving14} in the
generation of template waveforms for comparable mass compact binaries (made of
neutron star or black holes) to be analyzed in ground or space based
detectors. By contrast, self-force computations are designed with the view to
generate waveforms for comparison with the extreme mass ratio inspiral signals
expected from future space based detectors.
 
\begin{acknowledgments}
  We thank Sylvain Marsat for his independent proof of the basic
  formulas~\eqref{twosol}--\eqref{sol3}. This work was supported by
  NSF Grants PHY 1205906 and PHY 1314529 to UF. BFW acknowledges
  sabbatical support from the CNRS through the IAP, during the initial
  stages of this work, as well as support of the French state funds
  managed by the ANR within the Investissements d'Avenir programme
  under reference ANR-11-IDEX-0004-02. LB acknowledges hospitality and
  support from NSF and the Physics Department at UF, where part of
  this work was carried out during its final stages.
\end{acknowledgments}

\appendix

\section{Alternative computation of pole part contributions}
\label{app:alt}

According to our discussion in Section~\ref{sec:veq-ext}, in the coefficients given by
Eq.~\eqref{CBalt}, namely:
\begin{subequations}\label{CBgam}
\begin{align}
C_{k,\ell,m}(B) &=
\sum_{i=0}^\ell\gamma_{k,\ell,i}(B)\int_0^{+\infty}\ud
y\,y^i\frac{Q_m(1+y)}{(2+y)^{B-k+2}}\,,\label{CBexpr}\\ \text{where}
\quad\gamma_{k,\ell,i}(B) &= \frac{(\ell+i)!}{i!(\ell-i)!}
\frac{\Gamma(B-k-\ell+2)}{2^{i+1}\Gamma(B+1)}
\frac{\Gamma(B-k+\ell+3)}{\Gamma(B-k+i+3)} \,,\label{coeffgam}
\end{align}
\end{subequations}
we should compute the pole part when $B\to 0$.  Instead of expanding directly Eqs.~\eqref{CBgam} when $B \to 0$ (as
was done in Paper~I), it can be more convenient to expand an
equivalent, more explicit, expression of $C_{k,\ell,m}(B)$ that merely
differs from the original one by a finite remainder
$\mathcal{O}(B^0)$. This suggests an alternative method for computing the
poles at $B=0$, based on the fact that the integral
\begin{equation}\label{requform}
\mathcal{I}^\nu_{m} \equiv \int_0^{+\infty} \ud y \, y^\nu Q_m(1+y) =
2^{\nu} \frac{[\Gamma(\nu+1)]^2\,\Gamma(m-\nu)}{\Gamma(m+\nu+2)}\,,
\end{equation}
is known by analytic continuation for any $\nu\in\mathbb{C}$, except
at isolated poles at integer values of $\nu$ (see
\textit{e.g.}~\cite{GR}). The idea is to reshape the right-hand side
of Eq.~\eqref{CBexpr} in order to express it in terms of integrals
that possess the required form~\eqref{requform}. We proceed in two
steps.

(i) We perform the following transformation on the integrand of
Eq.~\eqref{CBexpr}. For $k\geqslant 2$, we write the denominator in
the original integrand as $(2+y)^{k-2}/(2+y)^B$ and expand
$(2+y)^{k-2}$ by means of the binomial theorem. This gives
($k\geqslant 2$)
\begin{equation}\label{cask2}
C_{k,\ell,m}(B) =
\sum_{i=0}^\ell\gamma_{k,\ell,i}(B)\sum_{j=i}^{i+k-2}2^{i+k-j-2}
    {\genfrac{(}{)}{0pt}{}{k-2}{j-i}}\int_0^{+\infty}\ud
    y\,y^j\,\frac{Q_m(1+y)}{(2+y)^{B}}\,,
\end{equation}
where ${\genfrac{(}{)}{0pt}{}{k-2}{j-i}}$ is the usual binomial
coefficient. For $k=1$, we replace the factor $y^i$ by the equivalent
form $(2+y) \sum_{j=0}^{i-1} (-2)^{i-j-1} y^j + (-2)^i$, which yields
\begin{equation}\label{cask10}
C_{1,\ell,m}(B) =
\sum_{i=0}^\ell\gamma_{1,\ell,i}(B)\sum_{j=0}^{i-1}(-2)^{i-j-1}
\int_0^{+\infty}\ud y\,y^j\,\frac{Q_m(1+y)}{(2+y)^{B}} +
\frac{1}{2}\int_0^{+\infty}\ud y\,\frac{Q_m(1+y)}{(2+y)^{B+1}}\,.
\end{equation}
The last integral in this expression corresponds to the contribution
to $C_{k,\ell, m}(B)$ produced by the term $(-2)^i$. Its coefficient has been
simplified by means of the following identity,
\begin{equation}\label{ident}
\sum_{i=0}^\ell(-2)^i\,\gamma_{1,\ell,i}(B) = \frac{1}{2}\,,
\end{equation}
resulting from the Gauss theorem on hypergeometric
functions. Remarkably, the last term in~\eqref{cask10} has no pole
at $B=0$ for $m \in \mathbb{N}$, since the integral is well defined in
the limit $B\to 0$, and it is therefore irrelevant for our
analysis. Thus we shall only need the expression
\begin{equation}\label{cask1}
C_{1,\ell,m}(B) =
\sum_{i=0}^\ell\gamma_{1,\ell,i}(B)\sum_{j=0}^{i-1}(-2)^{i-j-1}
\int_0^{+\infty}\ud y\,y^j\,\frac{Q_m(1+y)}{(2+y)^{B}} +
\mathcal{O}(B^0)\,.
\end{equation}
With Eqs.~\eqref{cask2} and~\eqref{cask1} in hands, we see that all
elementary integrands that may be associated with poles are now of the
type $y^j Q_m(1+y)/(2+y)^B$ (with $j \in \mathbb{N}$).

(ii) It is immediately possible to check that the pre-factors in
Eqs.~\eqref{cask2} and~\eqref{cask1} cannot have more than a simple
pole, so that it is sufficient to control the integrals of $y^j
Q_m(1+y)/(2+y)^B$ at order $\mathcal{O}(B^0)$, neglecting remainders
$\mathcal{O}(B)$. If $j<m$ the integral is convergent when $B=0$ and
its value is given by Eq.~\eqref{requform}. The problem is more
difficult when $j\geqslant m+1$. In that case we introduce the
asymptotic expansion of $y^j Q_m(1+y)$ when $y\to +\infty$. It is
obtained by expanding when $y\to +\infty$ the monomials, say
$(1+y)^{-2q-m-1}$ ($q\in \mathbb{N}$), in the hypergeometric series
defining $Q_m(1+y)$. After some technical manipulation involving again
the Gauss theorem, we get
\begin{subequations}\label{deffp}
\begin{align}
y^j Q_m(1+y) &= \sum_{p=0}^{j-m-1} f_{j,m,p}\,y^p +
\mathcal{O}\left(\frac{1}{y}\right)\,,\label{defpexp}\\ \text{where}
\quad f_{j,m,p} &=
(-)^{m}\frac{(-2)^{j-p-1}[(j-p-1)!]^2}{(j-m-p-1)!(j+m-p)!}\,.\label{fp}
\end{align}
\end{subequations}
Concretely, we shall resort to the following lemma, valid in the limit
$B\to 0$.
\begin{equation}\label{shiftint}
{\bf Lemma}:\quad\int_0^{+\infty}\ud y\,y^j\,\frac{Q_m(1+y)}{(2+y)^{B}} =
\int_0^{+\infty} \ud y\,y^{j-B}\,Q_m(1+y) + \sum_{p=0}^{j-m-1}
\frac{(-2)^{p+1}}{p+1} f_{j,m,p} + \mathcal{O}(B)\,.
\end{equation}
This permits us to relate the remaining integrals in~\eqref{cask2}
and~\eqref{cask1} to the simpler integrals that admit the closed-form
analytic expression~\eqref{requform}, with $\nu=j-B$. The proof relies on the
observation that, in the limit where $B\to 0$,
\begin{equation}\label{observ1}
\int_0^{+\infty}\ud
y\biggl(\frac{1}{(2+y)^B}-\frac{1}{y^B}\biggr)\biggl[y^j\,Q_m(1+y) -
  \sum_{p=0}^{j-m-1} f_{j,m,p}\,y^p\biggr] = \mathcal{O}(B)\,.
\end{equation}
This follows from the fact that the second factor inside the
integrand behaves like $\mathcal{O}(1/y)$ when $y\to +\infty$ [see
Eq.~\eqref{defpexp}]; so the integral is well-defined in a neighborhood
of $B=0$ and vanishes at that point. In addition, we can compute
explicitly, in the sense of analytic continuation in $B$ and in the
limit $B\to 0$, 
\begin{equation}\label{observ2}
\int_0^{+\infty}\ud y\,y^p\left(\frac{1}{(2+y)^B}-\frac{1}{y^B}\right)
= \frac{(-2)^{p+1}}{p+1} + \mathcal{O}(B)\,.
\end{equation}
The two facts~\eqref{observ1}--\eqref{observ2} imply
Eq.~\eqref{shiftint}.

Finally, transforming the integrals that enter Eqs.~\eqref{cask2}
and~\eqref{cask1} by means of our lemma~\eqref{shiftint}, when combined with
the expressions~\eqref{fp} for the coefficients $f_{j,m,p}$
and~\eqref{requform} for the integral $\mathcal{I}^{j-B}_{m}$, we obtain, in
the cases $k\geqslant 2$ and $k=1$ respectively,
\begin{subequations}\label{resultCklm}
\begin{align} 
C_{k,\ell,m}(B) &= 2^{k-3}(k-2)!  \frac{\Gamma(2-k-\ell+B) \Gamma(\ell +
  3-k+B)}{\Gamma(1+B)} \times \nonumber \\ & \quad ~ \times
\sum_{i=0}^{\ell+k-2} \frac{c_{k,\ell,i}(B)}{i!}  \bigg[
  \frac{[\Gamma(i+1-B)]^2 \Gamma(m-i+B)}{2^B \Gamma(m+i+2-B)} +
  (-)^{m+i}
  e_{i,m}\bigg]+\mathcal{O}(B^0)\,,\label{Cklm}\\ C_{1,\ell,m}(B) &=
\frac{\Gamma(1-\ell+B)\Gamma(\ell +2+B)}{4\Gamma(1+B)} \times
\nonumber \\ & \quad ~ \times \sum_{i=1}^{\ell} d_{\ell,i}(B) \bigg[
  \frac{[\Gamma(i-B)]^2\,\Gamma(m+1-i+B)}{2^B \Gamma(m+i+1-B)} +
  (-)^{m+i+1} e_{i-1,m} \bigg]+\mathcal{O}(B^0) \,.\label{C1lm}
\end{align}
\end{subequations}
The coefficients therein read
\begin{subequations}\label{coeffCklm}
\begin{align}
& c_{k,\ell,i}(B) = \sum_{j=\max(0,i+2-k)}^{\min(\ell,i)}
  {\genfrac{(}{)}{0pt}{}{i}{j}}\,\frac{(\ell+j)!}{(\ell-j)!(k+j-i-2)!}
  \frac{1}{\Gamma(j+3-k+B)} \, ,\\ & d_{\ell,i}(B) = \sum_{j=0}^{\ell-i}
  (-)^j \frac{(\ell+i+j)!}{(\ell-i-j)!(i+j)!}
  \frac{1}{\Gamma(i+j+2+B)} \, ,\\ & e_{i,m} = \sum_{j=0}^{i-m-1}
  \frac{[(i-j-1)!]^2}{(j+1)(i-j-m-1)! (m+i-j)!} \, .
\end{align}
\end{subequations}
The Laurent expansion when $B\to 0$ of the explicit
sums~\eqref{resultCklm}--\eqref{coeffCklm} can be performed rapidly in
a straightforward way.

\section{Source terms for the tails of tails}
\label{app:source}

The tail-of-tail terms associated with the various multipole moments
$I_L$ or $J_L$ (symbolized by $K_L$ say) obey a wave equation of the
type
\begin{equation}\label{Boxh3}
\Box h^{\alpha\beta}_{M \times M \times K_L} =
\Lambda^{\alpha\beta}_{M \times M \times K_L} \,,
\end{equation}
where $\Lambda_{M \times M \times K_L}$ is a cubic source term
composed of non-linear interactions between two static mass monopoles
$M$ and the time-varying multipole $K_L$. This source term has been
derived in Eqs.~(2.14)--(2.16) of Ref.~\cite{B98tail} for the
tails of tails associated with the mass quadrupole moment $I_{ij}$,
and this result was the basis of the computation of Paper~I. In this
Appendix we provide similar expressions for the sources of the
tails of tails associated with the mass moments $I_{ijk}$, $I_{ijkl}$
and current moments $J_{ij}$, $J_{ijk}$ that are also required for the
present computations. They have been obtained by means of the same
algorithm as in the quadrupolar case, using the \textit{xAct} package
bundle for Mathematica~\cite{xtensor}. As in Ref.~\cite{B98tail} and
Paper~I we split the source terms into an instantaneous
(local-in-time) part and a hereditary (past-dependent) one, say
\begin{equation}\label{calIH}
\Lambda^{\alpha\beta}_{M \times M \times K_L} =
\mathcal{I}^{\alpha\beta}_{M \times M \times K_L} +
\mathcal{H}^{\alpha\beta}_{M \times M \times K_L} \,.
\end{equation}
\begin{itemize}
\item Mass quadrupole moment:\footnote{We pose $G=c=1$ in this
  Appendix.}
\begin{subequations}\label{sourceinstI2}
\begin{align}
\mathcal{I}^{00}_{M \times M \times I_{ij}} &= M^2 n_{ab} r^{-7}
\biggl\{ -516 I_{ab} - 516 r I^{(1)}_{ab} - 304 r^2 I^{(2)}_{ab}
\nonumber\\&\qquad\qquad - 76 r^3 I^{(3)}_{ab} + 108 r^4 I^{(4)}_{ab}
+ 40 r^5 I^{(5)}_{ab} \biggr\} \,,\\
\mathcal{I}^{0i}_{M \times M \times I_{ij}} &= M^2 \hat{n}_{iab}
r^{-6} \biggl\{ 4 I^{(1)}_{ab} + 4 r I^{(2)}_{ab} - 16 r^2
I^{(3)}_{ab} + {4\over 3} r^3 I^{(4)}_{ab} - {4\over 3} r^4
I^{(5)}_{ab} \biggr\} \nonumber\\ &+ M^2 n_a r^{-6} \biggl\{
-{372\over 5} I^{(1)}_{ai} - {372\over 5} r I^{(2)}_{ai} -{232\over 5}
r^2 I^{(3)}_{ai} \nonumber\\ &\qquad\qquad- {84\over 5} r^3
I^{(4)}_{ai} + {124\over 5} r^4 I^{(5)}_{ai} \biggr\} \,, \\
\mathcal{I}^{ij}_{M \times M \times I_{ij}} &= M^2 \hat{n}_{ijab}
r^{-5} \biggl\{ -190 I^{(2)}_{ab} - 118 r I^{(3)}_{ab} - {92\over 3}
r^2 I^{(4)}_{ab} - 2 r^3 I^{(5)}_{ab} \biggr\} \nonumber\\&+
M^2\delta_{ij} n_{ab} r^{-5} \biggl\{ {160\over 7} I^{(2)}_{ab} +
{176\over 7} r I^{(3)}_{ab} - {596\over 21} r^2 I^{(4)}_{ab} -
{160\over 21} r^3 I^{(5)}_{ab} \biggr\} \nonumber\\ &+
M^2\hat{n}_{a(i} r^{-5} \biggl\{ -{312\over 7} I^{(2)}_{j)a} -
{248\over 7} r I^{(3)}_{j)a} + {400\over 7} r^2 I^{(4)}_{j)a} +
{104\over 7} r^3 I^{(5)}_{j)a} \biggr\} \nonumber\\ &+ M^2r^{-5}
\biggl\{ -12 I^{(2)}_{ij} - {196\over 15} r I^{(3)}_{ij} - {56\over 5}
r^2 I^{(4)}_{ij} - {48\over 5} r^3 I^{(5)}_{ij} \biggr\}\,.
\end{align}
\end{subequations}
\begin{subequations}\label{sourcetailI2}
\begin{align}
\mathcal{H}^{00}_{M \times M \times I_{ij}} &= M^2 n_{ab} r^{-3}
\int^{+\infty}_1 \ud x \biggl\{ 96 Q_0 I^{(4)}_{ab} + \left[ {272\over
    5} Q_1 + {168\over 5} Q_3 \right] r I^{(5)}_{ab} + 32 Q_2 r^2
I^{(6)}_{ab} \biggr\}\,,\\
\mathcal{H}^{0i}_{M \times M \times I_{ij}} &= M^2 \hat{n}_{iab} r^{-3}
\int^{+\infty}_1 \ud x \biggl\{ - 32 Q_1 I^{(4)}_{ab} + \left[
  -{32\over 3} Q_0 + {8\over 3} Q_2 \right] r I^{(5)}_{ab} \biggr\}
\nonumber\\ &+ M^2 n_a r^{-3} \int^{+\infty}_1 \ud x \biggl\{ {96\over
  5} Q_1 I^{(4)}_{ai} + \left[ {192\over 5} Q_0 + {112\over 5} Q_2
  \right] r I^{(5)}_{ai} + 32 Q_1 r^2 I^{(6)}_{ai} \biggr\}\,,\\
\mathcal{H}^{ij}_{M \times M \times I_{ij}} &= M^2 \hat{n}_{ijab}
r^{-3} \int^{+\infty}_1 \ud x \biggl\{ - 32 Q_2 I^{(4)}_{ab} + \left[-
  {32\over 5} Q_1 - {48\over 5} Q_3 \right] r I^{(5)}_{ab} \biggr\}
\nonumber\\ &+ M^2 \delta_{ij} n_{ab} r^{-3} \int^{+\infty}_1 \ud x
\biggl\{ - {32\over 7} Q_2 I^{(4)}_{ab} + \left[ - {208\over 7} Q_1 +
  {24\over 7} Q_3 \right] r I^{(5)}_{ab} \biggr\} \nonumber\\ &+ M^2
\hat{n}_{a(i} r^{-3} \int^{+\infty}_1 \ud x \biggl\{ {96\over 7} Q_2
I^{(4)}_{j)a} + \left[ {2112\over 35} Q_1 - {192\over 35} Q_3 \right]
r I^{(5)}_{j)a} \biggr\} \nonumber\\ &+ M^2 r^{-3} \int^{+\infty}_1
\ud x \biggl\{ {32\over 5} Q_2 I^{(4)}_{ij} + \left[ {512\over 25}
  Q_1 - {32\over 25} Q_3 \right] r I^{(5)}_{ij} + 32 Q_0 r^2
I^{(6)}_{ij} \biggr\}\,.
\end{align}
\end{subequations}
\item Mass octupole:
\begin{subequations}\label{sourceinstI3}
\begin{align}
\mathcal{I}^{00}_{M \times M \times I_{ijk}} &= M^2 \hat{n}_{abc}
r^{-8}\biggl\{ - 1140 I_{abc} - 1140 r I^{(1)}_{abc} - 616 r^{2}
I^{(2)}_{abc} - 236 r^{3} I^{(3)}_{abc} + \frac{76}{3} r^{4}
I^{(4)}_{abc} \nonumber \\ & \qquad \qquad \quad + \frac{484}{9} r^{5}
I_{abc}^{(5)} + \frac{112}{9} r^{6} I^{(6)}_{abc} \biggr\} \,,\\
\mathcal{I}^{0i}_{M \times M \times I_{ijk}} &=M^2 \hat{n}_{iabc}
r^{-7}\bigg\{ 6 I^{(1)}_{abc} +6 r I^{(2)}_{abc} - \frac{37}{3} r^2
I^{(3)}_{abc} - \frac{43}{3} r^3 I^{(4)}_{abc} - \frac{16}{9} r^4
I^{(5)}_{abc} - \frac{1}{3} r^5 I^{(6)}_{abc} \bigg\} \nonumber \\ & +
M^2 \hat{n}_{ab} r^{-7}\bigg\{- \frac{892}{7} I^{(1)}_{abi} -
\frac{892}{7} r I^{(2)}_{abi} - \frac{492}{7} r^2 I^{(3)}_{abi} -
\frac{584}{21} r^3 I^{(4)}_{abi} \nonumber \\ & \qquad \qquad \quad +
\frac{568}{63} r^4 I^{(5)}_{abi} + \frac{572}{63} r^5I^{(6)}_{abi}
\bigg\} \,, \\
\mathcal{I}^{ij}_{M \times M \times I_{ijk}} &= M^2 \hat{n}_{ijabc}
r^{-6} \bigg\{ -186 I^{(2)}_{abc} -186 r I^{(3)}_{abc} - 68 r^2
I^{(4)}_{abc} - \frac{34}{3} r^3 I^{(5)}_{abc} - \frac{8}{15} r^4
I^{(6)}_{abc} \bigg\} \nonumber \\ & + M^2 \delta_{ij} \hat{n}_{abc}
r^{-6} \bigg\{ 24 I^{(2)}_{abc} + 24 r I^{(3)}_{abc} - \frac{38}{9}
r^2 I^{(4)}_{abc} - \frac{346}{27} r^3 I^{(5)}_{abc} - \frac{46}{27}
r^4 I^{(6)}_{abc} \bigg\} \nonumber \\ & + M^2 \hat{n}_{ab(i} r^{-6}
\bigg\{ - \frac{140}{3} I^{(2)}_{j)ab} - \frac{140}{3} r
I^{(3)}_{j)ab} + \frac{38}{3} r^2 I^{(4)}_{j)ab} \nonumber \\ & \qquad 
\qquad \qquad + \frac{230}{9} r^3
I^{(5)}_{j)ab} + \frac{10}{3} r^{4}I^{(6)}_{j)ab} \bigg\} \nonumber
\\ & + M^2 n_{a} r^{-6} \bigg\{- \frac{356}{21} I^{(2)}_{aij}-
\frac{356}{21} r I^{(3)}_{aij} - \frac{1028}{105} r^2 I^{(4)}_{aij} -
\frac{296}{63} r^3 I^{(5)}_{aij} + \frac{24}{7} r^4 I^{(6)}_{aij}
\bigg\} \,.
\end{align}
\end{subequations}
\begin{subequations} \label{sourcetailI3}
\begin{align}
\mathcal{H}^{00}_{M \times M \times I_{ijk}} &= 
M^2 \hat{n}_{abc} r^{-3} \int_{1}^{+\infty} \ud x \bigg\{ 
32  I^{(5)}_{abc} Q_{1} + 
\frac{8}{3}  \biggl[ 7 Q_{2} + 4 Q_{4}\biggr] r I^{(6)}_{abc} + 
 \frac{32}{3} Q_{3} r^{2} I^{(7)}_{abc}  \bigg\} \, , \\
\mathcal{H}^{0i}_{M \times M \times I_{ijk}} &= M^2 \hat{n}_{iabc}
r^{-3}\int_{1}^{+\infty} \ud x \bigg\{- \frac{64}{3} Q_{2}
I^{(5)}_{abc} - \frac{8}{15} \bigg[8 Q_{1} - 3 Q_{3}\bigg] r
I^{(6)}_{abc} \bigg\} \nonumber \\ & + M^2 \hat{n}_{ab} r^{-3}
\int_{1}^{+\infty} \ud x \bigg\{ \frac{256}{21} Q_{2} I^{(5)}_{abi}
\nonumber \\ & \qquad \qquad \qquad \qquad \quad + \frac{8}{105}
\bigg[172 Q_{1} + 93 Q_{3}\bigg] r I^{(6)}_{abi} + \frac{32}{3} Q_{2}
r^2 I^{(7)}_{abi} \bigg\} \, , \\
\mathcal{H}^{ij}_{M \times M \times I_{ijk}} &= M^2 \hat{n}_{ijabc}
r^{-3} \int_{1}^{+\infty} \ud x \bigg\{ -16 Q_{3} I^{(5)}_{abc} -
\frac{16}{21} \bigg[ 3 Q_{2} + 4 Q_{4} \bigg] r I^{(6)}_{abc} \bigg\}
\nonumber \\ & + M^2 \delta_{ij} \hat{n}_{abc} r^{-3}
\int_{1}^{+\infty} \ud x \bigg\{- \frac{16}{9} Q_{3} I^{(5)}_{abc} -
\frac{8}{27} \bigg[33 Q_{2} - 4 Q_{4}\bigg] r I^{(6)}_{abc} \bigg\}
\nonumber \\ & + M^2 \hat{n}_{ab(i} r^{-3} \int_{1}^{+\infty} \ud x
\bigg\{ \frac{16}{3} Q_{3} I^{(5)}_{j)ab} + \frac{32}{63} \bigg[39
  Q_{2} - 4 Q_{4}\bigg] r I^{(6)}_{j)ab} \bigg\} \nonumber \\ & + M^2
n_{a} r^{-3} \int_{1}^{+\infty} \ud x \bigg\{ \frac{128}{35} Q_{3}
I^{(5)}_{aij} + \frac{32}{735} \bigg[187 Q_{2} - 12 Q_{4}\bigg] r
I^{(6)}_{aij} \nonumber \\ & \qquad \qquad \qquad \qquad \quad + 
\frac{32}{3} Q_{1} r^2 I^{(7)}_{aij} \bigg\} \, .
\end{align}
\end{subequations}
\item Mass hexadecapole:
\begin{subequations} \label{sourceinstI4}
\begin{align}
\mathcal{I}^{00}_{M \times M \times I_{ijkl}}&= M^2 \hat{n}_{abcd}
r^{-9} \bigg\{ - 2520 I_{abcd}- 2520 r I^{(1)}_{abcd} - 1318 r^2
I^{(2)}_{abcd}- 478 r^3 I^{(3)}_{abcd} \nonumber \\ & \qquad \qquad \quad -
\frac{1015}{18} r^4 I^{(4)}_{abcd}+ \frac{845}{18} r^5 I^{(5)}_{abcd}
+ \frac{133}{6} r^6 I^{(6)}_{abcd} + \frac{29}{9} r^7 I^{(7)}_{abcd}
\bigg\}\, , \\
\mathcal{I}^{0i}_{M \times M \times I_{ijkl}}&=  M^2 \hat{n}_{iabcd} r^{-8}
\bigg\{ 9 I^{(1)}_{abcd} + 9 r I^{(2)}_{abcd} - 
\frac{29}{2} r^2 I^{(3)}_{abcd} - 
\frac{35}{2} r^3 I^{(4)}_{abcd} - 
\frac{68}{9} r^4 I^{(5)}_{abcd} \nonumber \\ & \qquad \qquad \qquad -
\frac{77}{90} r^5 I^{(6)}_{abcd} - 
\frac{1}{15} r^6 I^{(7)}_{abcd} \bigg\} \nonumber \\ & +
M^2 \hat{n}_{abc} r^{-8} \bigg\{ - 230  I^{(1)}_{abci} -  
230 r I^{(2)}_{abci} -  \frac{1093}{9} r^2 I^{(3)}_{abci} -  
\frac{403}{9} r^3 I^{(4)}_{abci}  \nonumber \\ & \qquad \qquad \quad -
\frac{119}{81} r^4 I^{(5)}_{abci} +
\frac{646}{81} r^5 I^{(6)}_{abci} +
 \frac{70}{27} r^6 I^{(7)}_{abci} \bigg\} \, , \\
\mathcal{I}^{ij}_{M \times M \times I_{ijkl}}&= M^2 \hat{n}_{ijabcd}
r^{-7} \bigg\{-293 I^{(2)}_{abcd} - 293 r I^{(3)}_{abcd} -
\frac{379}{3} r^2 I^{(4)}_{abcd} \nonumber \\ & \qquad \qquad \qquad -
\frac{86}{3} r^3 I^{(5)}_{abcd} - \frac{146}{45} r^4 I^{(6)}_{abcd} -
\frac{1}{9} r^5 I^{(7)}_{abcd} \bigg\} \nonumber \\ & + M^2
\delta_{ij} \hat{n}_{abcd} r^{-7} \bigg\{ \frac{345}{11}
I^{(2)}_{abcd} + \frac{345}{11} r I^{(3)}_{abcd} + \frac{355}{198} r^2
I^{(4)}_{abcd} - \frac{1715}{198} r^3 I^{(5)}_{abcd} \nonumber \\ &
\qquad \qquad \qquad - \frac{4087}{990} r^4 I^{(6)}_{abcd}-
\frac{53}{165} r^5 I^{(7)}_{abcd} \bigg\} \nonumber \\ & + M^2
\hat{n}_{abc(i} r^{-7} \bigg\{ - \frac{672}{11} I^{(2)}_{j)abc} -
\frac{672}{11} r I^{(3)}_{j)abc} - \frac{208}{99} r^2 I^{(4)}_{j)abc}
+ \frac{1808}{99} r^3 I^{(5)}_{j)abc} \nonumber \\ & \qquad \qquad
\quad + \frac{452}{55} r^4 I^{(6)}_{j)abc} + \frac{104}{165} r^5
I^{(7)}_{j)abc} \bigg\} \nonumber \\ & + M^2 \hat{n}_{ab}
r^{-7}\bigg\{ - \frac{74}{3} I^{(2)}_{abij} - \frac{74}{3} r
I^{(3)}_{abij} - \frac{835}{63} r^2 I^{(4)}_{abij} - \frac{317}{63}
r^3 I^{(5)}_{abij} \nonumber \\ & \qquad \qquad \quad +
\frac{110}{189} r^4I^{(6)}_{abij} + \frac{314}{189} r^5 I^{(7)}_{abij}
\bigg\} \, .
\end{align}
\end{subequations}
\begin{subequations} \label{sourcetailI4}
\begin{align}
\mathcal{H}^{00}_{M \times M \times I_{ijkl}}&=
M^2 \hat{n}_{abcd} r^{-3} \int_{1}^{+\infty} \ud x
\bigg\{8  Q_{2} I^{(6)}_{abcd} + 
\frac{2}{27} \bigg[64 Q_{3}+ 35 Q_{5} \bigg] r I^{(7)}_{abcd}  \nonumber \\ &
\qquad \qquad \qquad \qquad \quad  + 
\frac{8}{3} Q_{4} r^2 I^{(8)}_{abcd} \bigg\} \, , \\
\mathcal{H}^{0i}_{M \times M \times I_{ijkl}}&= M^2 \hat{n}_{iabcd}
r^{-3} \int_{1}^{+\infty} \ud x \bigg\{- 8 Q_{3} I^{(6)}_{abcd} -
\frac{2}{21} \bigg[ 12 Q_{2}- 5 Q_{4} \bigg] r I^{(7)}_{abcd} \bigg\}
\nonumber \\ & + M^2 \hat{n}_{abc} r^{-3} \int_{1}^{+\infty} \ud x
\bigg\{ \frac{40}{9} Q_{3} I^{(6)}_{abci} \nonumber \\ & \qquad \qquad
\qquad \qquad \quad + \frac{8}{189} \bigg[78 Q_{2} + 41 Q_{4}\bigg] r
I^{(7)}_{abci} + \frac{8}{3} Q_{3} r^2 I^{(8)}_{abci}\bigg\} \, , \\
\mathcal{H}^{ij}_{M \times M \times I_{ijkl}}&= M^2 \hat{n}_{ijabcd}
r^{-3} \int_{1}^{+\infty} \ud x \bigg\{ - \frac{16}{3} Q_{4}
I^{(6)}_{abcd} - \frac{4}{27} \bigg[4 Q_{3} + 5 Q_{5}\bigg] r
I^{(7)}_{abcd} \bigg\} \nonumber \\ & + M^2 \delta_{ij} \hat{n}_{abcd}
r^{-3} \int_{1}^{+\infty} \ud x \bigg\{ - \frac{16}{33} Q_{4}
I^{(6)}_{abcd} - \frac{10}{33} \bigg[8 Q_{3} - Q_{5}\bigg] r
I^{(7)}_{abcd} \bigg\} \nonumber \\ & + M^2 \hat{n}_{abc(i} r^{-3}
\int_{1}^{+\infty} \ud x \bigg\{ \frac{16}{11} Q_{4} I^{(6)}_{j)abc} +
\frac{16}{297} \bigg[91 Q_{3} - 10 Q_{5}\bigg] r I^{(7)}_{j)abc}
\bigg\} \nonumber \\ & + M^2 \hat{n}_{ab} r^{-3} \int_{1}^{+\infty}
\ud x \bigg\{ \frac{80}{63} Q_{4} I^{(6)}_{abij} + \frac{16}{567}
\bigg[77 Q_{3} - 5 Q_{5}\bigg] r I^{(7)}_{abij} \nonumber \\ & \qquad \qquad
\qquad \qquad \quad + \frac{8}{3} Q_{2}
r^2 I^{(8)}_{abij} \bigg\} \, .
\end{align}
\end{subequations}
\item Current quadrupole:
\begin{subequations} \label{sourceinstJ2}
\begin{align}
\mathcal{I}^{00}_{M \times M \times J_{ij}}&=0 \, , \\
\mathcal{I}^{0i}_{M \times M \times J_{ij}}&= M^2 \varepsilon_{iab}
\hat{n}_{ac} r^{-7} \bigg\{ 88 J_{bc} + 88 r J^{(1)}_{bc}+ 80 r^2
J^{(2)}_{bc} + \frac{152}{3} r^3 J^{(3)}_{bc} \nonumber \\ & \qquad
\qquad \qquad - \frac{368}{9} r^4 J^{(4)}_{bc} - \frac{208}{9} r^5
J^{(5)}_{bc} \bigg\} \, , \\
\mathcal{I}^{ij}_{M \times M \times J_{ij}}&= M^2 \varepsilon_{ab(i}
\hat{n}_{j)ac} r^{-6} \bigg\{ \frac{64}{3} J^{(1)}_{bc} +\frac{64}{3}
r J^{(2)}_{bc} -64 r^2 J^{(3)}_{bc} - \frac{608}{9} r^3 J^{(4)}_{bc} -
\frac{64}{9} r^4 J^{(5)}_{bc} \bigg\} \nonumber \\ & + M^2
\varepsilon_{ab(i} n_{\underline{a}} r^{-6} \bigg\{\frac{304}{15}
J^{(1)}_{j)b} + \frac{304}{15} r J^{(2)}_{j)b}+ \frac{368}{15} r^2
J^{(3)}_{j)b} \nonumber \\ & \qquad \qquad \qquad + \frac{832}{45} r^3
J^{(4)}_{j)b}- \frac{96}{5} r^4 J^{(5)}_{j)b} \bigg\}\, .
\end{align}
\end{subequations}
\begin{subequations} \label{sourcetailJ2}
\begin{align}
\mathcal{H}^{00}_{M \times M \times J_{ij}}&=0 \, , \\
\mathcal{H}^{0i}_{M \times M \times J_{ij}}&= M^2
\varepsilon_{iab}\hat{n}_{ac} r^{-3}\int_{1}^{+\infty} \ud x \bigg\{
\frac{64}{3} Q_{2} J^{(4)}_{bc} - \frac{64}{15} \bigg[7 Q_{1} + 3
  Q_{3} \bigg] r J^{(5)}_{bc} \nonumber \\ & \qquad \qquad \qquad 
\qquad \qquad - \frac{64}{3} Q_{2} r^2 J^{(6)}_{bc}
\bigg\} \, , \\
\mathcal{H}^{ij}_{M \times M \times J_{ij}}&=M^2 \varepsilon_{ab(i}
\hat{n}_{j)ac} r^{-2} \int_{1}^{+\infty} \ud x  \bigg\{
-  \frac{128}{3}  Q_{2} J^{(5)}_{bc}  \bigg\} \nonumber
\\ & +  M^2 \varepsilon_{ab(i} n_{\underline{a}}   
r^{-2}\int_{1}^{+\infty} \ud x\bigg\{- \frac{128}{15} Q_{2} J^{(5)}_{j)b} - 
\frac{128}{3} Q_{1} r J^{(6)}_{j)b} \bigg\} \, .
\end{align}
\end{subequations}
\item Current octupole:
\begin{subequations} \label{sourceinstJ3}
\begin{align}
\mathcal{I}^{00}_{M \times M \times J_{ijk}} &=0 \, , \\
\mathcal{I}^{0i}_{M \times M \times J_{ijk}}&=M^2 \varepsilon_{iab}
\hat{n}_{acd} r^{-8} \bigg\{ 270 J_{bcd} + 270 r J^{(1)}_{bcd} + 188
r^2 J^{(2)}_{bcd} + 98 r^3 J^{(3)}_{bcd} - 6 r^4 J^{(4)}_{bcd}
\nonumber \\ & \qquad \qquad \qquad \quad - \frac{100}{3} r^5
J^{(5)}_{bcd} - 9 r^6 J^{(6)}_{bcd} \bigg\} \, , \\
\mathcal{I}^{ij}_{M \times M \times J_{ijk}}&=M^2 \varepsilon_{ab(i}
\hat{n}_{j)acd} r^{-7} \bigg\{ 54 J^{(1)}_{bcd} + 54 r J^{(2)}_{bcd} -
60 r^2 J^{(3)}_{bcd}-78 r^3 J^{(4)}_{bcd} \nonumber \\ & \qquad \qquad
\qquad \quad - 30 r^4 J^{(5)}_{bcd}- 2 r^5 J^{(6)}_{bcd}\bigg\}
\nonumber \\ & +M^2 \varepsilon_{ab(i}
\hat{n}_{\underline{a}\underline{c}} r^{-7} \bigg\{ \frac{360}{7}
J^{(1)}_{j)bc} + \frac{360}{7} r J^{(2)}_{j)bc} + \frac{286}{7} r^2
J^{(3)}_{j)bc} + \frac{166}{7} r^3 J^{(4)}_{j)bc} \nonumber \\ &
\qquad \qquad \qquad \quad - \frac{32}{7} r^4 J^{(5)}_{j)bc}-
\frac{236}{21}r^5 J^{(6)}_{j)bc} \bigg\} \, .
\end{align}
\end{subequations}
\begin{subequations} \label{sourcetailJ3}
\begin{align}
\mathcal{H}^{00}_{M \times M \times J_{ijk}} &=0 \, , \\
\mathcal{H}^{0i}_{M \times M \times J_{ijk}}&=M^2 \varepsilon_{iab}
\hat{n}_{acd}  r^{-3} \int_{1}^{+\infty} \ud x\bigg\{8 Q_{3} J^{(5)}_{bcd} -   
\frac{16}{7} \bigg[5 Q_{2} + 2 Q_{4}\bigg]  r J^{(6)}_{bcd} -
8 Q_{3} r^2  J^{(7)}_{bcd} \bigg\} \, , \\
\mathcal{H}^{ij}_{M \times M \times J_{ijk}}&= M^2 \varepsilon_{ab(i}
\hat{n}_{j)acd} r^{-2} \int_{1}^{+\infty} \ud x \bigg\{ -16 Q_{3}
J^{(6)}_{bcd} \bigg\} \nonumber \\ &+ M^2 \varepsilon_{ab(i}
\hat{n}_{\underline{a}\underline{c}} r^{-2} \int_{1}^{+\infty} \ud x
\bigg\{- \frac{32}{7} Q_{3} J^{(6)}_{j)bc} - 16 Q_{2} r J^{(7)}_{j)bc}
\bigg\} \, .
\end{align}
\end{subequations}
\end{itemize}
In the hereditary terms the kernels of the integrals are made of
Legendre functions of the second kind $Q_m(x)$, see Eq.~\eqref{Qm},
multiplied by time derivatives of the multipole moments evaluated at
time $t-r x$.

In the Appendix~A of Paper~I, it was proved that certain specific 
terms, namely those coming from the second term in
Eq.~\eqref{formuleZP}, do not contribute at half-integral
PN orders.  It is easy to verify that the proof there applies in the more 
general case investigated here, where we have additional multipole 
components besides the mass quadrupole.  
Indeed we observe that for all the
hereditary terms in Eqs.~\eqref{sourceinstI2}--\eqref{sourcetailJ3}
the combination $k+m+\ell$ is always an odd integer, where $k$
represents the power of $1/r$ in the term in question, $m$ is the
order of the Legendre function therein and $\ell$ is the multipolarity
of the term. Thus the proof of Appendix~A in Paper~I can be repeated
exactly as it is. This shows that the PN order of the second term in
Eq.~\eqref{formuleZP}, for all these multipole interactions, is
necessarily integral and can be ignored in the present computation.

\bibliography{ListeRef.bib}

\begin{thebibliography}{49}
\expandafter\ifx\csname natexlab\endcsname\relax\def\natexlab#1{#1}\fi
\expandafter\ifx\csname bibnamefont\endcsname\relax
  \def\bibnamefont#1{#1}\fi
\expandafter\ifx\csname bibfnamefont\endcsname\relax
  \def\bibfnamefont#1{#1}\fi
\expandafter\ifx\csname citenamefont\endcsname\relax
  \def\citenamefont#1{#1}\fi
\expandafter\ifx\csname url\endcsname\relax
  \def\url#1{\texttt{#1}}\fi
\expandafter\ifx\csname urlprefix\endcsname\relax\def\urlprefix{URL }\fi
\providecommand{\bibinfo}[2]{#2}
\providecommand{\eprint}[2][]{\url{#2}}

\bibitem[{\citenamefont{Detweiler}(2008)}]{Det08}
\bibinfo{author}{\bibfnamefont{S.}~\bibnamefont{Detweiler}},
  \bibinfo{journal}{Phys. Rev. D} \textbf{\bibinfo{volume}{77}},
  \bibinfo{pages}{124026} (\bibinfo{year}{2008}), \eprint{arXiv:0804.3529
  [gr-qc]}.

\bibitem[{\citenamefont{Barack and Sago}(2009)}]{BarackS09}
\bibinfo{author}{\bibfnamefont{L.}~\bibnamefont{Barack}} \bibnamefont{and}
  \bibinfo{author}{\bibfnamefont{N.}~\bibnamefont{Sago}},
  \bibinfo{journal}{Phys. Rev. Lett.} \textbf{\bibinfo{volume}{102}},
  \bibinfo{pages}{191101} (\bibinfo{year}{2009}), \eprint{arXiv:0902.0573
  [gr-qc]}.

\bibitem[{\citenamefont{Blanchet
  et~al.}(2010{\natexlab{a}})\citenamefont{Blanchet, Detweiler, Le~Tiec, and
  Whiting}}]{BDLW10a}
\bibinfo{author}{\bibfnamefont{L.}~\bibnamefont{Blanchet}},
  \bibinfo{author}{\bibfnamefont{S.}~\bibnamefont{Detweiler}},
  \bibinfo{author}{\bibfnamefont{A.}~\bibnamefont{Le~Tiec}}, \bibnamefont{and}
  \bibinfo{author}{\bibfnamefont{B.~F.} \bibnamefont{Whiting}},
  \bibinfo{journal}{Phys. Rev. D} \textbf{\bibinfo{volume}{81}},
  \bibinfo{pages}{064004} (\bibinfo{year}{2010}{\natexlab{a}}),
  \eprint{arXiv:0910.0207 [gr-qc]}.

\bibitem[{\citenamefont{Blanchet
  et~al.}(2010{\natexlab{b}})\citenamefont{Blanchet, Detweiler, Le~Tiec, and
  Whiting}}]{BDLW10b}
\bibinfo{author}{\bibfnamefont{L.}~\bibnamefont{Blanchet}},
  \bibinfo{author}{\bibfnamefont{S.}~\bibnamefont{Detweiler}},
  \bibinfo{author}{\bibfnamefont{A.}~\bibnamefont{Le~Tiec}}, \bibnamefont{and}
  \bibinfo{author}{\bibfnamefont{B.~F.} \bibnamefont{Whiting}},
  \bibinfo{journal}{Phys. Rev. D} \textbf{\bibinfo{volume}{81}},
  \bibinfo{pages}{084033} (\bibinfo{year}{2010}{\natexlab{b}}),
  \eprint{arXiv:1002.0726 [gr-qc]}.

\bibitem[{\citenamefont{Keidl et~al.}(2010)\citenamefont{Keidl, Shah, Friedman,
  Kim, and Price}}]{Keidl10}
\bibinfo{author}{\bibfnamefont{T.~S.} \bibnamefont{Keidl}},
  \bibinfo{author}{\bibfnamefont{A.~G.} \bibnamefont{Shah}},
  \bibinfo{author}{\bibfnamefont{J.~L.} \bibnamefont{Friedman}},
  \bibinfo{author}{\bibfnamefont{D.-H.} \bibnamefont{Kim}}, \bibnamefont{and}
  \bibinfo{author}{\bibfnamefont{L.~R.} \bibnamefont{Price}},
  \bibinfo{journal}{Phys. Rev. D} \textbf{\bibinfo{volume}{82}},
  \bibinfo{pages}{124012} (\bibinfo{year}{2010}).

\bibitem[{\citenamefont{Shah et~al.}(2011)\citenamefont{Shah, Keidl, Friedman,
  Kim, and Price}}]{Shah11}
\bibinfo{author}{\bibfnamefont{A.~G.} \bibnamefont{Shah}},
  \bibinfo{author}{\bibfnamefont{T.~S.} \bibnamefont{Keidl}},
  \bibinfo{author}{\bibfnamefont{J.~L.} \bibnamefont{Friedman}},
  \bibinfo{author}{\bibfnamefont{D.-H.} \bibnamefont{Kim}}, \bibnamefont{and}
  \bibinfo{author}{\bibfnamefont{L.~R.} \bibnamefont{Price}},
  \bibinfo{journal}{Phys. Rev. D} \textbf{\bibinfo{volume}{83}},
  \bibinfo{pages}{064018} (\bibinfo{year}{2011}).

\bibitem[{\citenamefont{Shah et~al.}(2012)\citenamefont{Shah, Friedman, and
  Keidl}}]{Shah12}
\bibinfo{author}{\bibfnamefont{A.~G.} \bibnamefont{Shah}},
  \bibinfo{author}{\bibfnamefont{J.~L.} \bibnamefont{Friedman}},
  \bibnamefont{and} \bibinfo{author}{\bibfnamefont{T.~S.} \bibnamefont{Keidl}},
  \bibinfo{journal}{Phys. Rev. D} \textbf{\bibinfo{volume}{86}},
  \bibinfo{pages}{084059} (\bibinfo{year}{2012}).

\bibitem[{\citenamefont{Akcay et~al.}(2012)\citenamefont{Akcay, Barack, Damour,
  and Sago}}]{Akcay12}
\bibinfo{author}{\bibfnamefont{S.}~\bibnamefont{Akcay}},
  \bibinfo{author}{\bibfnamefont{L.}~\bibnamefont{Barack}},
  \bibinfo{author}{\bibfnamefont{T.}~\bibnamefont{Damour}}, \bibnamefont{and}
  \bibinfo{author}{\bibfnamefont{N.}~\bibnamefont{Sago}},
  \bibinfo{journal}{Phys. Rev. D} \textbf{\bibinfo{volume}{86}},
  \bibinfo{pages}{104041} (\bibinfo{year}{2012}).

\bibitem[{\citenamefont{Teukolsky}(1972)}]{Teukolsky72}
\bibinfo{author}{\bibfnamefont{S.~A.} \bibnamefont{Teukolsky}},
  \bibinfo{journal}{Phys. Rev. Lett.} \textbf{\bibinfo{volume}{29}},
  \bibinfo{pages}{1114} (\bibinfo{year}{1972}).

\bibitem[{\citenamefont{Teukolsky}(1973)}]{Teukolsky73}
\bibinfo{author}{\bibfnamefont{S.~A.} \bibnamefont{Teukolsky}},
  \bibinfo{journal}{Astrophys. J.} \textbf{\bibinfo{volume}{185}},
  \bibinfo{pages}{635} (\bibinfo{year}{1973}).

\bibitem[{\citenamefont{Keidl et~al.}(2007)\citenamefont{Keidl, Friedman, and
  Wiseman}}]{Keidl07}
\bibinfo{author}{\bibfnamefont{T.~S.} \bibnamefont{Keidl}},
  \bibinfo{author}{\bibfnamefont{J.~L.} \bibnamefont{Friedman}},
  \bibnamefont{and} \bibinfo{author}{\bibfnamefont{A.~G.}
  \bibnamefont{Wiseman}}, \bibinfo{journal}{Phys. Rev. D}
  \textbf{\bibinfo{volume}{75}}, \bibinfo{pages}{124009}
  (\bibinfo{year}{2007}).

\bibitem[{\citenamefont{Regge and Wheeler}(1957)}]{ReggeW57}
\bibinfo{author}{\bibfnamefont{T.}~\bibnamefont{Regge}} \bibnamefont{and}
  \bibinfo{author}{\bibfnamefont{J.~A.} \bibnamefont{Wheeler}},
  \bibinfo{journal}{Phys. Rev.} \textbf{\bibinfo{volume}{108}},
  \bibinfo{pages}{1063} (\bibinfo{year}{1957}).

\bibitem[{\citenamefont{Zerilli}(1970)}]{Ze70}
\bibinfo{author}{\bibfnamefont{F.~J.} \bibnamefont{Zerilli}},
  \bibinfo{journal}{Phys. Rev. Lett.} \textbf{\bibinfo{volume}{24}},
  \bibinfo{pages}{737} (\bibinfo{year}{1970}).

\bibitem[{\citenamefont{Le~Tiec
  et~al.}(2012{\natexlab{a}})\citenamefont{Le~Tiec, Blanchet, and
  Whiting}}]{LBW12}
\bibinfo{author}{\bibfnamefont{A.}~\bibnamefont{Le~Tiec}},
  \bibinfo{author}{\bibfnamefont{L.}~\bibnamefont{Blanchet}}, \bibnamefont{and}
  \bibinfo{author}{\bibfnamefont{B.~F.} \bibnamefont{Whiting}},
  \bibinfo{journal}{Phys. Rev. D} \textbf{\bibinfo{volume}{85}},
  \bibinfo{pages}{064039} (\bibinfo{year}{2012}{\natexlab{a}}),
  \eprint{arXiv:1111.5378 [gr-qc]}.

\bibitem[{\citenamefont{Shah et~al.}(2014)\citenamefont{Shah, Friedmann, and
  Whiting}}]{SFW14}
\bibinfo{author}{\bibfnamefont{A.~G.} \bibnamefont{Shah}},
  \bibinfo{author}{\bibfnamefont{J.~L.} \bibnamefont{Friedmann}},
  \bibnamefont{and} \bibinfo{author}{\bibfnamefont{B.~F.}
  \bibnamefont{Whiting}}, \bibinfo{journal}{Phys. Rev. D}
  \textbf{\bibinfo{volume}{89}}, \bibinfo{pages}{064042}
  (\bibinfo{year}{2014}), \eprint{arXiv:1312.1952 [gr-qc]}.

\bibitem[{\citenamefont{Bini and Damour}(2013)}]{BiniD13}
\bibinfo{author}{\bibfnamefont{D.}~\bibnamefont{Bini}} \bibnamefont{and}
  \bibinfo{author}{\bibfnamefont{T.}~\bibnamefont{Damour}},
  \bibinfo{journal}{Phys. Rev. D} \textbf{\bibinfo{volume}{87}},
  \bibinfo{pages}{121501(R)} (\bibinfo{year}{2013}), \eprint{arXiv:1305.4884
  [gr-qc]}.

\bibitem[{\citenamefont{Bini and Damour}(2014{\natexlab{a}})}]{BiniD14a}
\bibinfo{author}{\bibfnamefont{D.}~\bibnamefont{Bini}} \bibnamefont{and}
  \bibinfo{author}{\bibfnamefont{T.}~\bibnamefont{Damour}},
  \bibinfo{journal}{Phys. Rev. D} \textbf{\bibinfo{volume}{89}},
  \bibinfo{pages}{064063} (\bibinfo{year}{2014}{\natexlab{a}}),
  \eprint{arXiv:1312.2503 [gr-qc]}.

\bibitem[{\citenamefont{Bini and Damour}(2014{\natexlab{b}})}]{BiniD14b}
\bibinfo{author}{\bibfnamefont{D.}~\bibnamefont{Bini}} \bibnamefont{and}
  \bibinfo{author}{\bibfnamefont{T.}~\bibnamefont{Damour}},
  \bibinfo{journal}{Phys. Rev. D} \textbf{\bibinfo{volume}{89}},
  \bibinfo{pages}{104047} (\bibinfo{year}{2014}{\natexlab{b}}),
  \eprint{arXiv:1403.2366 [gr-qc]}.

\bibitem[{\citenamefont{Mano et~al.}(1996{\natexlab{a}})\citenamefont{Mano,
  Susuki, and Takasugi}}]{MST96a}
\bibinfo{author}{\bibfnamefont{S.}~\bibnamefont{Mano}},
  \bibinfo{author}{\bibfnamefont{H.}~\bibnamefont{Susuki}}, \bibnamefont{and}
  \bibinfo{author}{\bibfnamefont{E.}~\bibnamefont{Takasugi}},
  \bibinfo{journal}{Prog. Theor. Phys.} \textbf{\bibinfo{volume}{95}},
  \bibinfo{pages}{1079} (\bibinfo{year}{1996}{\natexlab{a}}),
  \eprint{gr-qc/9603020}.

\bibitem[{\citenamefont{Mano et~al.}(1996{\natexlab{b}})\citenamefont{Mano,
  Susuki, and Takasugi}}]{MST96b}
\bibinfo{author}{\bibfnamefont{S.}~\bibnamefont{Mano}},
  \bibinfo{author}{\bibfnamefont{H.}~\bibnamefont{Susuki}}, \bibnamefont{and}
  \bibinfo{author}{\bibfnamefont{E.}~\bibnamefont{Takasugi}},
  \bibinfo{journal}{Prog. Theor. Phys.} \textbf{\bibinfo{volume}{96}},
  \bibinfo{pages}{549} (\bibinfo{year}{1996}{\natexlab{b}}),
  \eprint{gr-qc/9605057}.

\bibitem[{\citenamefont{Mano and Takasugi}(1997)}]{MT97}
\bibinfo{author}{\bibfnamefont{S.}~\bibnamefont{Mano}} \bibnamefont{and}
  \bibinfo{author}{\bibfnamefont{E.}~\bibnamefont{Takasugi}},
  \bibinfo{journal}{Prog. Theor. Phys.} \textbf{\bibinfo{volume}{97}},
  \bibinfo{pages}{213} (\bibinfo{year}{1997}), \eprint{gr-qc/9611014}.

\bibitem[{\citenamefont{Blanchet et~al.}(2014)\citenamefont{Blanchet, Faye, and
  Whiting}}]{BFW14a}
\bibinfo{author}{\bibfnamefont{L.}~\bibnamefont{Blanchet}},
  \bibinfo{author}{\bibfnamefont{G.}~\bibnamefont{Faye}}, \bibnamefont{and}
  \bibinfo{author}{\bibfnamefont{B.~F.} \bibnamefont{Whiting}},
  \bibinfo{journal}{Phys. Rev. D} \textbf{\bibinfo{volume}{89}},
  \bibinfo{pages}{064026} (\bibinfo{year}{2014}), \eprint{arXiv:1312.2975
  [gr-qc]}.

\bibitem[{\citenamefont{Mino et~al.}(1997)\citenamefont{Mino, Sasaki, and
  Tanaka}}]{MiSaTa}
\bibinfo{author}{\bibfnamefont{Y.}~\bibnamefont{Mino}},
  \bibinfo{author}{\bibfnamefont{M.}~\bibnamefont{Sasaki}}, \bibnamefont{and}
  \bibinfo{author}{\bibfnamefont{T.}~\bibnamefont{Tanaka}},
  \bibinfo{journal}{Phys. Rev. D} \textbf{\bibinfo{volume}{55}},
  \bibinfo{pages}{3457} (\bibinfo{year}{1997}), \eprint{gr-qc/9606018}.

\bibitem[{\citenamefont{Quinn and Wald}(1997)}]{QuWa}
\bibinfo{author}{\bibfnamefont{T.~C.} \bibnamefont{Quinn}} \bibnamefont{and}
  \bibinfo{author}{\bibfnamefont{R.~M.} \bibnamefont{Wald}},
  \bibinfo{journal}{Phys. Rev. D} \textbf{\bibinfo{volume}{56}},
  \bibinfo{pages}{3381} (\bibinfo{year}{1997}), \eprint{gr-qc/9610053}.

\bibitem[{\citenamefont{Detweiler and Whiting}(2003)}]{DW03}
\bibinfo{author}{\bibfnamefont{S.}~\bibnamefont{Detweiler}} \bibnamefont{and}
  \bibinfo{author}{\bibfnamefont{B.~F.} \bibnamefont{Whiting}},
  \bibinfo{journal}{Phys. Rev. D} \textbf{\bibinfo{volume}{67}},
  \bibinfo{pages}{024025} (\bibinfo{year}{2003}), \eprint{gr-qc/0202086}.

\bibitem[{\citenamefont{Gralla and Wald}(2008)}]{GW08}
\bibinfo{author}{\bibfnamefont{S.}~\bibnamefont{Gralla}} \bibnamefont{and}
  \bibinfo{author}{\bibfnamefont{R.}~\bibnamefont{Wald}},
  \bibinfo{journal}{Class. Quant. Grav.} \textbf{\bibinfo{volume}{25}},
  \bibinfo{pages}{205009} (\bibinfo{year}{2008}), \eprint{arXiv:0806.3293
  [gr-qc]}.

\bibitem[{\citenamefont{Pound}(2010)}]{Pound10}
\bibinfo{author}{\bibfnamefont{A.}~\bibnamefont{Pound}},
  \bibinfo{journal}{Phys. Rev. D} \textbf{\bibinfo{volume}{81}},
  \bibinfo{pages}{024023} (\bibinfo{year}{2010}), \eprint{arXiv:0907.5197
  [gr-qc]}.

\bibitem[{\citenamefont{Poisson et~al.}(2011)\citenamefont{Poisson, Pound, and
  Vega}}]{PoissonLR}
\bibinfo{author}{\bibfnamefont{E.}~\bibnamefont{Poisson}},
  \bibinfo{author}{\bibfnamefont{A.}~\bibnamefont{Pound}}, \bibnamefont{and}
  \bibinfo{author}{\bibfnamefont{I.}~\bibnamefont{Vega}},
  \bibinfo{journal}{Living Rev. Rel.} \textbf{\bibinfo{volume}{14}},
  \bibinfo{pages}{7} (\bibinfo{year}{2011}), \eprint{arXiv:1102.0529 [gr-qc]}.

\bibitem[{\citenamefont{Detweiler}(2011)}]{Detweilerorleans}
\bibinfo{author}{\bibfnamefont{S.}~\bibnamefont{Detweiler}}, in
  \emph{\bibinfo{booktitle}{Mass and motion in general relativity}}, edited by
  \bibinfo{editor}{\bibfnamefont{L.}~\bibnamefont{Blanchet}},
  \bibinfo{editor}{\bibfnamefont{A.}~\bibnamefont{Spallicci}},
  \bibnamefont{and} \bibinfo{editor}{\bibfnamefont{B.}~\bibnamefont{Whiting}}
  (\bibinfo{publisher}{Springer}, \bibinfo{year}{2011}), p.
  \bibinfo{pages}{271}.

\bibitem[{\citenamefont{Barack}(2011)}]{Barackorleans}
\bibinfo{author}{\bibfnamefont{L.}~\bibnamefont{Barack}}, in
  \emph{\bibinfo{booktitle}{Mass and motion in general relativity}}, edited by
  \bibinfo{editor}{\bibfnamefont{L.}~\bibnamefont{Blanchet}},
  \bibinfo{editor}{\bibfnamefont{A.}~\bibnamefont{Spallicci}},
  \bibnamefont{and} \bibinfo{editor}{\bibfnamefont{B.}~\bibnamefont{Whiting}}
  (\bibinfo{publisher}{Springer}, \bibinfo{year}{2011}), p.
  \bibinfo{pages}{327}.

\bibitem[{\citenamefont{De~Witt and Brehme}(1960)}]{dWB60}
\bibinfo{author}{\bibfnamefont{B.}~\bibnamefont{De~Witt}} \bibnamefont{and}
  \bibinfo{author}{\bibfnamefont{R.}~\bibnamefont{Brehme}},
  \bibinfo{journal}{Ann. Phys. (N.Y.)} \textbf{\bibinfo{volume}{9}},
  \bibinfo{pages}{220} (\bibinfo{year}{1960}).

\bibitem[{\citenamefont{Blanchet}(2014)}]{Bliving14}
\bibinfo{author}{\bibfnamefont{L.}~\bibnamefont{Blanchet}},
  \bibinfo{journal}{Living Rev. Rel.} \textbf{\bibinfo{volume}{17}},
  \bibinfo{pages}{2} (\bibinfo{year}{2014}), \eprint{arXiv:1310.1528 [gr-qc]}.

\bibitem[{\citenamefont{Blanchet et~al.}(1998)\citenamefont{Blanchet, Faye, and
  Ponsot}}]{BFP98}
\bibinfo{author}{\bibfnamefont{L.}~\bibnamefont{Blanchet}},
  \bibinfo{author}{\bibfnamefont{G.}~\bibnamefont{Faye}}, \bibnamefont{and}
  \bibinfo{author}{\bibfnamefont{B.}~\bibnamefont{Ponsot}},
  \bibinfo{journal}{Phys. Rev. D} \textbf{\bibinfo{volume}{58}},
  \bibinfo{pages}{124002} (\bibinfo{year}{1998}), \eprint{gr-qc/9804079}.

\bibitem[{\citenamefont{Jaranowski and Sch{\"a}fer}(2012)}]{JaraS12}
\bibinfo{author}{\bibfnamefont{P.}~\bibnamefont{Jaranowski}} \bibnamefont{and}
  \bibinfo{author}{\bibfnamefont{G.}~\bibnamefont{Sch{\"a}fer}},
  \bibinfo{journal}{Phys. Rev. D} \textbf{\bibinfo{volume}{86}},
  \bibinfo{pages}{061503(R)} (\bibinfo{year}{2012}), \eprint{arXiv:1207.5448
  [gr-qc]}.

\bibitem[{\citenamefont{Jaranowski and Sch{\"a}fer}(2013)}]{JaraS13}
\bibinfo{author}{\bibfnamefont{P.}~\bibnamefont{Jaranowski}} \bibnamefont{and}
  \bibinfo{author}{\bibfnamefont{G.}~\bibnamefont{Sch{\"a}fer}},
  \bibinfo{journal}{Phys. Rev. D} \textbf{\bibinfo{volume}{87}},
  \bibinfo{pages}{081503(R)} (\bibinfo{year}{2013}), \eprint{arXiv:1303.3225
  [gr-qc]}.

\bibitem[{\citenamefont{Blanchet}(1998{\natexlab{a}})}]{B98tail}
\bibinfo{author}{\bibfnamefont{L.}~\bibnamefont{Blanchet}},
  \bibinfo{journal}{Class. Quant. Grav.} \textbf{\bibinfo{volume}{15}},
  \bibinfo{pages}{113} (\bibinfo{year}{1998}{\natexlab{a}}),
  \bibinfo{note}{erratum Class. Quant. Grav. {\bf 22}, 3381 (2005)},
  \eprint{gr-qc/9710038}.

\bibitem[{\citenamefont{Le~Tiec
  et~al.}(2012{\natexlab{b}})\citenamefont{Le~Tiec, Barausse, and
  Buonanno}}]{LBB12}
\bibinfo{author}{\bibfnamefont{A.}~\bibnamefont{Le~Tiec}},
  \bibinfo{author}{\bibfnamefont{E.}~\bibnamefont{Barausse}}, \bibnamefont{and}
  \bibinfo{author}{\bibfnamefont{A.}~\bibnamefont{Buonanno}},
  \bibinfo{journal}{Phys. Rev. Lett.} \textbf{\bibinfo{volume}{108}},
  \bibinfo{pages}{131103} (\bibinfo{year}{2012}{\natexlab{b}}),
  \eprint{arXiv:1111.5609 [gr-qc]}.

\bibitem[{\citenamefont{Barausse et~al.}(2012)\citenamefont{Barausse, Buonanno,
  and Le~Tiec}}]{BBL11}
\bibinfo{author}{\bibfnamefont{E.}~\bibnamefont{Barausse}},
  \bibinfo{author}{\bibfnamefont{A.}~\bibnamefont{Buonanno}}, \bibnamefont{and}
  \bibinfo{author}{\bibfnamefont{A.}~\bibnamefont{Le~Tiec}},
  \bibinfo{journal}{Phys.Rev. D} \textbf{\bibinfo{volume}{85}},
  \bibinfo{pages}{064010} (\bibinfo{year}{2012}), \eprint{arXiv:1111.5610
  [gr-qc]}.

\bibitem[{\citenamefont{Blanchet and Damour}(1986)}]{BD86}
\bibinfo{author}{\bibfnamefont{L.}~\bibnamefont{Blanchet}} \bibnamefont{and}
  \bibinfo{author}{\bibfnamefont{T.}~\bibnamefont{Damour}},
  \bibinfo{journal}{Phil. Trans. Roy. Soc. Lond. A}
  \textbf{\bibinfo{volume}{320}}, \bibinfo{pages}{379} (\bibinfo{year}{1986}).

\bibitem[{\citenamefont{Blanchet}(1998{\natexlab{b}})}]{B98quad}
\bibinfo{author}{\bibfnamefont{L.}~\bibnamefont{Blanchet}},
  \bibinfo{journal}{Class. Quant. Grav.} \textbf{\bibinfo{volume}{15}},
  \bibinfo{pages}{89} (\bibinfo{year}{1998}{\natexlab{b}}),
  \eprint{gr-qc/9710037}.

\bibitem[{\citenamefont{Blanchet}(1993)}]{B93}
\bibinfo{author}{\bibfnamefont{L.}~\bibnamefont{Blanchet}},
  \bibinfo{journal}{Phys. Rev. D} \textbf{\bibinfo{volume}{47}},
  \bibinfo{pages}{4392} (\bibinfo{year}{1993}).

\bibitem[{\citenamefont{Burke and Thorne}(1970)}]{BuTh70}
\bibinfo{author}{\bibfnamefont{W.}~\bibnamefont{Burke}} \bibnamefont{and}
  \bibinfo{author}{\bibfnamefont{K.}~\bibnamefont{Thorne}}, in
  \emph{\bibinfo{booktitle}{Relativity}}, edited by
  \bibinfo{editor}{\bibfnamefont{M.}~\bibnamefont{Carmeli}},
  \bibinfo{editor}{\bibfnamefont{S.}~\bibnamefont{Fickler}}, \bibnamefont{and}
  \bibinfo{editor}{\bibfnamefont{L.}~\bibnamefont{Witten}}
  (\bibinfo{publisher}{Plenum Press}, \bibinfo{address}{New York and London},
  \bibinfo{year}{1970}), pp. \bibinfo{pages}{209--228}.

\bibitem[{\citenamefont{Burke}(1971)}]{Bu71}
\bibinfo{author}{\bibfnamefont{W.}~\bibnamefont{Burke}}, \bibinfo{journal}{J.
  Math. Phys.} \textbf{\bibinfo{volume}{12}}, \bibinfo{pages}{401}
  (\bibinfo{year}{1971}).

\bibitem[{\citenamefont{Misner et~al.}(1973)\citenamefont{Misner, Thorne, and
  Wheeler}}]{MTW}
\bibinfo{author}{\bibfnamefont{C.}~\bibnamefont{Misner}},
  \bibinfo{author}{\bibfnamefont{K.}~\bibnamefont{Thorne}}, \bibnamefont{and}
  \bibinfo{author}{\bibfnamefont{J.}~\bibnamefont{Wheeler}},
  \emph{\bibinfo{title}{Gravitation}} (\bibinfo{publisher}{Freeman},
  \bibinfo{address}{San Francisco}, \bibinfo{year}{1973}).

\bibitem[{\citenamefont{Fock}(1959)}]{Fock}
\bibinfo{author}{\bibfnamefont{V.}~\bibnamefont{Fock}},
  \emph{\bibinfo{title}{Theory of space, time and gravitation}}
  (\bibinfo{publisher}{Pergamon}, \bibinfo{address}{London},
  \bibinfo{year}{1959}).

\bibitem[{\citenamefont{Schwartz}(1978)}]{Schwartz}
\bibinfo{author}{\bibfnamefont{L.}~\bibnamefont{Schwartz}},
  \emph{\bibinfo{title}{Th\'eorie des distributions}}
  (\bibinfo{publisher}{Hermann}, \bibinfo{address}{Paris},
  \bibinfo{year}{1978}).

\bibitem[{\citenamefont{Blanchet and Faye}(2001)}]{BFeom}
\bibinfo{author}{\bibfnamefont{L.}~\bibnamefont{Blanchet}} \bibnamefont{and}
  \bibinfo{author}{\bibfnamefont{G.}~\bibnamefont{Faye}},
  \bibinfo{journal}{Phys. Rev. D} \textbf{\bibinfo{volume}{63}},
  \bibinfo{pages}{062005} (\bibinfo{year}{2001}), \eprint{gr-qc/0007051}.

\bibitem[{\citenamefont{Gradshteyn and Ryzhik}(1980)}]{GR}
\bibinfo{author}{\bibfnamefont{I.}~\bibnamefont{Gradshteyn}} \bibnamefont{and}
  \bibinfo{author}{\bibfnamefont{I.}~\bibnamefont{Ryzhik}},
  \emph{\bibinfo{title}{Table of Integrals, Series and Products}}
  (\bibinfo{publisher}{Academic Press}, \bibinfo{year}{1980}).

\bibitem[{\citenamefont{Mart\'in-Garc\'ia et~al.}(GPL
  2002--2012)\citenamefont{Mart\'in-Garc\'ia, Garc\'ia-Parrado, Stecchina,
  Wardell, Pitrou, Brizuela, Yllanes, Faye, Stein, Portugal et~al.}}]{xtensor}
\bibinfo{author}{\bibfnamefont{J.~M.} \bibnamefont{Mart\'in-Garc\'ia}},
  \bibinfo{author}{\bibfnamefont{A.}~\bibnamefont{Garc\'ia-Parrado}},
  \bibinfo{author}{\bibfnamefont{A.}~\bibnamefont{Stecchina}},
  \bibinfo{author}{\bibfnamefont{B.}~\bibnamefont{Wardell}},
  \bibinfo{author}{\bibfnamefont{C.}~\bibnamefont{Pitrou}},
  \bibinfo{author}{\bibfnamefont{D.}~\bibnamefont{Brizuela}},
  \bibinfo{author}{\bibfnamefont{D.}~\bibnamefont{Yllanes}},
  \bibinfo{author}{\bibfnamefont{G.}~\bibnamefont{Faye}},
  \bibinfo{author}{\bibfnamefont{L.}~\bibnamefont{Stein}},
  \bibinfo{author}{\bibfnamefont{R.}~\bibnamefont{Portugal}},
  \bibnamefont{et~al.}, \emph{\bibinfo{title}{{xAct}: Efficient tensor computer
  algebra for {Mathematica}}} (\bibinfo{year}{GPL 2002--2012}),
  \bibinfo{note}{http://www.xact.es/}.

\end{thebibliography}

\end{document}